# Cooperative Cognitive Networks: Optimal, Distributed and Low-Complexity Algorithms

Gan Zheng[*], S. H. Song[†], Kai-Kit Wong[‡], and Björn Ottersten[*]

**Abstract**

This paper considers the cooperation between a cognitive system and a primary system where multiple cognitive base stations (CBSs) relay the primary user's (PU) signals in exchange for more opportunity to transmit their own signals. The CBSs use amplify-and-forward (AF) relaying and coordinated beamforming to relay the primary signals and transmit their own signals. The objective is to minimize the overall transmit power of the CBSs given the rate requirements of the PU and the cognitive users (CUs). We show that the relaying matrices have unit rank and perform two functions: Matched filter receive beamforming and transmit beamforming. We then develop two efficient algorithms to find the optimal solution. The first one has linear convergence rate and is suitable for distributed implementation, while the second one enjoys superlinear convergence but requires centralized processing. Further, we derive the beamforming vectors for the linear conventional zero-forcing (CZF) and prior zero-forcing (PZF) schemes, which provide much simpler solutions. Simulation results demonstrate the improvement in terms of outage performance due to the cooperation between the primary and cognitive systems.

[*]Interdisciplinary Centre for Security, Reliability and Trust (SnT), University of Luxembourg, L-1359 Luxembourg-Kirchberg, Luxembourg. Email: {gan.zheng, bjorn.ottersten}@uni.lu. B. Ottersten is also with the Signal Processing Laboratory, ACCESS Linnaeus Center, KTH Royal Institute of Technology, Sweden. Email: bjorn.ottersten@ee.kth.se.

[†]Department of Electronic and Computer Engineering, The Hong Kong University of Science and Technology, Clear Water Bay, Kowloon, Hong Kong. Email: shsong@ieee.org.

[‡]Department of Electronic and Electrical Engineering, University College London, UK. Email: kai-kit.wong@ucl.ac.uk.

# I. Introduction

Communications over wireless channels continue to be major challenges of today's technologies mainly due to spectrum scarcity and channel fading characteristics. While spectrum utilization depends very much on place and time, it has been well known that most spectrum is heavily under-utilized [1]. Cognitive radio system (CRS) [2] is a new paradigm to improve the spectrum efficiency by allowing a secondary system to access the spectrum licensed to the primary system. In a typical setup, the primary users (PUs) have the priority to access the spectrum, while CRS can occupy the spectrum only if it does not interrupt the communication of the primary system. In practice, this either requires the CRS to sense and detect the spectrum holes and then access the spectrum opportunistically or demands the interference from CRS to the primary system to be properly controlled. In either case, the primary system and CRS are working separately. A major challenge therefore arises to guarantee the quality-of-service (QoS) of cognitive users (CUs) without degrading the primary system performance. In light of this, a number of beamforming techniques have been proposed to achieve various related objectives assuming perfect [3], partial [4] or imperfect channel state information (CSI) [5, 6] available at the CRS regarding the primary system.

On the other hand, cooperative communication, especially via relaying, is a promising countermeasure for channel fading. Relaying strategies may be categorized into three main types: 1) compress-and-forward (CF) 2) amplify-and-forward (AF) and 3) decode-and-forward (DF). Among them, AF, in which the relay simply performs linear processing on the received noisy signal from the sender and forwards it to the destination, is arguably the most attractive strategy, due to its relatively low implementation complexity. Interested readers are referred to the special issues such as [7] for more details on this topic.

Cooperation can be used to improve the reliability performance of CRS as well as the primary system. The advantages of cooperation are manifold. It can help the CRS forward the primary and cognitive signals [8], spectrum sensing [9, 10] and save energy in the process of cooperative sensing and transmission [11]. Without introducing additional relay nodes, the CRS itself can serve as a relay. Cognitive relaying was proposed in multiple access control (MAC) layer [12] where CRS directly helps to relay the traffic of the primary system. The benefit is that the PUs' performance can be improved due to the diversity paths via CBS and in return, CRS gets a higher chance to utilize the spectrum, hence increasing the overall spectral efficiency. This is particularly important to the primary system when the PUs' instantaneous data rate or outage probability cannot be satisfied, due to a weak PU link. For this reason, the primary system is incentivised to provide necessary CSI to the CRS for cooperation.

Multiple antennas at the CBS can also greatly benefit the cognitive relaying, but it is relatively unexplored. In [14], cognitive beamforming was devised to broadcast a common message to the CUs in the downlink using semidefinite programming (SDP) relaxation. Nevertheless, in the case of AF relaying, the results in [14] are limited because CBS only uses a vector to linearly process and forward the signal from the primary base station (PBS), and thus is unable to fully exploit the multiple antennas at the CBS. In [13], a DF relaying protocol was considered for the downlink and prior zero-forcing (PZF), which gives priority to the primary system, was proposed and compared with the conventional zero-forcing (CZF) algorithm. A rate threshold was also given as a guideline on which method is preferred. Despite these early works, joint optimal relay processing and transmit beamforming is unknown even with one CBS.

## A. Our Aims and Contributions

This paper studies the downlink scenario where several multi-antenna CBSs help to relay the PU signal using AF strategy while serving the CUs.[1] Our aim is to minimize the total CBS transmit power (for both relaying and own transmission) for given PU and CU target rates by optimizing the relaying matrices and the coordinated transmit beamforming vectors [15]. Regarding the requirement of CSI, PBS does not need to learn its channels to CRS because its transmission strategy is unchanged. It is the CBSs who should acquire the global CSI, including CSI from the PBS, to the PU and to all CUs. The first is received CSI and can be estimated by training; the second can also be obtained either via the help of PBS or by listening to the feedback from the PU during the channel estimation of the primary system, and this is possible since in order to cooperate, CBSs should not be too far from the PBS. For the third channels from one CBS to all CUs, they can be learned by feedback from the CUs involved. In summary, perfect, either local or global, CSI is assumed at the CBSs. We have made the following contributions:

- We consider cooperation among multiple CBSs using coordinated beamforming in addition to the cooperation between CBS and the primary system for enhancing the overall spectrum efficiency. The joint design of coordinated beamforming *and* collaborative cognitive relaying is original and novel.
- The optimal structure of the relaying matrices is shown to match the backward channels for maximizing the received signal-to-noise ratio (SNR) of the PU signal and align the noisy PU signal collaboratively using relay beamforming. This structure greatly simplifies the original problem.

[1]As a possible application, the cooperative cognitive system will be useful to repair a satellite link which suffers from high path loss and rain fading. For instance, several terrestrial base stations can form a cognitive network and use the same spectrum to serve their own users, while at the same time help relay the satellite signal from the ground station. This improves the availability and saves on board satellite power, which is of great importance for extending satellite lifetime.

- Two iterative methods are devised to achieve the optimal solution to the total power minimization problem. The first one is based on a fixed-point iteration, has linear convergence, and permits a distributed implementation that requires only *local* CSI at the CBS while another method utilizes matrix iteration, has superlinear convergence but requires *global* CSI at the CBS.

- In addition, we derive closed-form expressions for the beamforming vectors for the CZF and PZF schemes, which provide linear and simple solutions. A unique rate requirement threshold is given to differentiate the advantages between CZF and PZF for the case with only one CBS and one CU.

## B. Notations

Throughout this paper, the following notations will be adopted. Vectors and matrices are represented by boldface lowercase and uppercase letters, respectively. $\|\cdot\|_x$ is the $l_x$ norm and $\|\cdot\|$ denotes the Frobenius norm. $\mathsf{vec}(\mathbf{x}_1,\dots,\mathbf{x}_M)$ returns a column vector by stacking all the elements of the input vectors $\mathbf{x}_1,\dots,\mathbf{x}_M$ in order. $(\cdot)^\dagger$ denotes the Hermitian operation of a vector or matrix, while $\mathsf{trace}(\mathbf{A})$ returns the trace of $\mathbf{A}$. The notation $\mathbf{A}\in\mathbb{R}^{M\times N}$ indicates that $\mathbf{A}$ is a real matrix with dimension $M\times N$. $\mathbf{A}\succeq\mathbf{0}$ means that $\mathbf{A}$ is positive semi-definite. In addition, $\mathbf{A}\propto\mathbf{B}$ means that $\mathbf{A}=c\mathbf{B}$ where $c$ is a constant. $\mathbf{A}_{m,n}$ denotes the $(m,n)$-th element of $\mathbf{A}$. Also, $\mathsf{BlkDiag}(\mathbf{A}_1,\dots,\mathbf{A}_M)$ returns a block diagonal matrix with $\mathbf{A}_1,\dots,\mathbf{A}_M$ being diagonal matrices. $\mathbf{I}$ denotes an identity matrix of appropriate sizes. Finally, $\mathbf{x}\sim\mathcal{CN}(\mathbf{m},\mathbf{\Theta})$ denotes a vector $\mathbf{x}$ of complex Gaussian entries with a mean vector of $\mathbf{m}$ and a covariance matrix of $\mathbf{\Theta}$.

# II. Network Model and Problem Formulation

## A. The Primary and Cognitive System

In this paper, we consider a primary-secondary downlink network as shown in Fig. 1. The primary system consists of a $K$-antenna PBS communicating to a single-antenna PU, whereas the secondary system has $M$ CBSs each with $N$ transmit/receive antennas and serving a single-antenna CU.[2] Without loss of generality, it is assumed that the $m$-th CU and its serving CBS are denoted by $\mathsf{CU}_m$ and $\mathsf{CBS}_m$, respectively. Of our interest is the scenario where cooperation between CBSs and PBS is necessary to support an acceptable QoS of the primary system. Various possible working modes of cooperation will be discussed in Section

[2] The extension to the case that one CBS serves multiple CUs is straightforward.

II-C. To describe our network model, we define

| | |
|---|---|
| $\mathbf{h}_0^{p\dagger}$ | the $1 \times K$ channel vector between the PBS and the PU; |
| $\mathbf{h}_{m0}^{\dagger}$ | the $1 \times N$ channel vector between $\mathsf{CBS}_m$ and the PU; |
| $\mathbf{h}_{jm}^{\dagger}$ | the $1 \times N$ channel vector between $\mathsf{CBS}_j$ and $\mathsf{CU}_m$; |
| $\mathbf{G}_m$ | the channel between the PBS and $\mathsf{CBS}_m$; |
| $n_1^p$ | the noise received at the PU during phase I with $n_1^p \sim \mathcal{CN}(0, N_1^p)$; |
| $n_2^p$ | the noise received at the PU during phase II with $n_2^p \sim \mathcal{CN}(0, N_2^p)$; |
| $\mathbf{n}_{sm}$ | the $N \times 1$ noise vector received at $\mathsf{CBS}_m$ during phase I with $\mathbf{n}_{sm} \sim \mathcal{CN}(\mathbf{0}, N_{sm}\mathbf{I})$; |
| $n_m$ | the noise received at $\mathsf{CU}_m$ during phase II with $n_m \sim \mathcal{CN}(0, N_m)$; |
| $P_0$ | the transmit power of PBS; |
| $s_0$ | the transmit signal for the PU with $s_0 \sim \mathcal{CN}(0, P_0)$; |
| $s_m$ | the transmit signal for $\mathsf{CU}_m$ with $s_m \sim \mathcal{CN}(0, 1)$. |

All channels and noises are independent of each other, and the CBSs operate in an AF and half-duplex fashion. The communication is synchronous and takes place in two phases. In phase I, PBS broadcasts its message $s_0$ by $\mathbf{x} = \mathbf{f}s_0$ where $\mathbf{f}$ is a fixed unit-norm transmit beamforming vector which is either $\frac{\mathbf{h}_0^p}{\|\mathbf{h}_0^p\|}$ or to be designed by the PBS. The received signals at the PU and $\mathsf{CBS}_m$ are, respectively, given by

$$y_1 = \mathbf{h}_0^{p\dagger}\mathbf{f}s_0 + n_1^p, \tag{1}$$

$$\mathbf{r}_m = \mathbf{G}_m\mathbf{f}s_0 + \mathbf{n}_{sm} = \mathbf{g}_m s_0 + \mathbf{n}_{sm}, \tag{2}$$

where $\mathbf{g}_m \triangleq \mathbf{G}_m\mathbf{f}, \forall m$. Then, $\mathsf{CBS}_m$ processes the received signal using an $N \times N$ relaying matrix $\mathbf{A}_m$ to produce $\mathbf{A}_m\mathbf{r}_m = \mathbf{A}_m\mathbf{g}_m s_0 + \mathbf{A}_m\mathbf{n}_{sm}$. In phase II, $\mathsf{CBS}_m$ sends its own message, $s_m$, using a beamforming vector $\{\mathbf{w}_m\}$ together with the processed signals $\mathbf{A}_m\mathbf{r}_m$. We assume that data symbols for the CUs, $\{s_m\}$, and the PU, $s_0$, are mutually independent of each other. During this period, the PBS remains idle and the received signal at $\mathsf{CU}_m$ is then given by

$$z_m = \mathbf{h}_{mm}^{\dagger}\left(\mathbf{w}_m s_m + \mathbf{A}_m\mathbf{g}_m s_0 + \mathbf{A}_m\mathbf{n}_{sm}\right) + \sum_{\substack{j=1\\ j\neq m}}^{M} \mathbf{h}_{jm}^{\dagger}\left(\mathbf{w}_j s_j + \mathbf{A}_j\mathbf{g}_j s_0 + \mathbf{A}_j\mathbf{n}_{sj}\right) + n_m. \tag{3}$$

The received signal-to-interference plus noise ratio (SINR) at $\mathsf{CU}_m$ is then expressed as

$$\mathsf{SINR}|_{\mathsf{CU}_m} = \frac{|\mathbf{h}_{mm}^{\dagger}\mathbf{w}_m|^2}{\sum_{\substack{j=1\\j\neq m}}^{M} |\mathbf{h}_{jm}^{\dagger}\mathbf{w}_j|^2 + P_0 \sum_{j=1}^{M} |\mathbf{h}_{jm}^{\dagger}\mathbf{A}_j\mathbf{g}_j|^2 + \sum_{j=1}^{M} N_{sj}\|\mathbf{h}_{jm}^{\dagger}\mathbf{A}_j\|^2 + N_m}. \tag{4}$$

Similarly, the received signal at the PU is given by

$$\begin{aligned} y_2 &= \sum_{j=1}^{M} \mathbf{h}_{j0}^{\dagger}\left(\mathbf{w}_j s_j + \mathbf{A}_j\mathbf{g}_j s_0 + \mathbf{A}_j\mathbf{n}_{sj}\right) + n_2^p \\ &= \sum_{j=1}^{M} \mathbf{h}_{j0}^{\dagger}\mathbf{A}_j\mathbf{g}_j s_0 + \sum_{j=1}^{M} \mathbf{h}_{j0}^{\dagger}\mathbf{w}_j s_j + \sum_{j=1}^{M} \mathbf{h}_{j0}^{\dagger}\mathbf{A}_j\mathbf{n}_{sj} + n_2^p. \end{aligned} \tag{5}$$

Using maximal-ratio combining (MRC) to get an estimate of $s_0$, the received SINR of the PU becomes

$$\mathsf{SINR}|_{\mathsf{PU}} = P_0\left(\frac{|\mathbf{h}_0^{p\dagger}\mathbf{f}|^2}{N_1^p} + \frac{|\sum_{j=1}^{M}\mathbf{h}_{j0}^{\dagger}\mathbf{A}_j\mathbf{g}_j|^2}{\sum_{j=1}^{M}|\mathbf{h}_{j0}^{\dagger}\mathbf{w}_j|^2 + \sum_{j=1}^{M} N_{sj}\|\mathbf{h}_{j0}^{\dagger}\mathbf{A}_j\|^2 + N_2^p}\right). \tag{6}$$

## B. Problem Formulation

Assuming perfect CSI, we aim to minimize the transmit power of CBSs subject to both PU and CUs' rate constraints $r_0$ and $\{r_m\}$, respectively, by jointly optimizing $\{\mathbf{w}_m\}$ and $\{\mathbf{A}_m\}$. Mathematically, that is,

$$\begin{aligned} \min_{\{\mathbf{w}_j,\mathbf{A}_j\}} \quad & \sum_{j=1}^{M}\|\mathbf{w}_j\|^2 + P_0\sum_{j=1}^{M}\|\mathbf{A}_j\mathbf{g}_j\|^2 + \sum_{j=1}^{M} N_{sj}\|\mathbf{A}_j\|^2 \\ \text{s.t.} \quad & \frac{|\sum_{j=1}^{M}\mathbf{h}_{j0}^{\dagger}\mathbf{A}_j\mathbf{g}_j|^2}{\sum_{j=1}^{M}|\mathbf{h}_{j0}^{\dagger}\mathbf{w}_j|^2 + \sum_{j=1}^{M} N_{sj}\|\mathbf{h}_{j0}^{\dagger}\mathbf{A}_j\|^2 + N_2^p} \geq \gamma_0^{'} \triangleq \frac{2^{2r_0}-1}{P_0} - \frac{|\mathbf{h}_0^{p\dagger}\mathbf{f}|^2}{N_1^p}, \\ & \frac{|\mathbf{h}_{mm}^{\dagger}\mathbf{w}_m|^2}{\sum_{\substack{j=1\\j\neq m}}^{M}|\mathbf{h}_{jm}^{\dagger}\mathbf{w}_j|^2 + P_0\sum_{j=1}^{M}|\mathbf{h}_{jm}^{\dagger}\mathbf{A}_j\mathbf{g}_j|^2 + \sum_{j=1}^{M} N_{sj}\|\mathbf{h}_{jm}^{\dagger}\mathbf{A}_j\|^2 + N_m} \geq \gamma_m \triangleq 2^{2r_m}-1, \forall m. \end{aligned} \tag{7}$$

In this paper, we mainly consider the case that the original primary link $\mathbf{h}_0^p$ is weak and cooperation will be beneficial to the primary system. The above formulation has assumed that $r_0$ is greater than the rate that is achievable by the primary link alone. Therefore, it is necessary to use cooperation to improve the spectral efficiency for the PU. To understand this, the achievable rate for the PU depends on: 1) PU transmit power $P_0$, 2) the channel quality of PBS-CBS link $\mathbf{g}_j$ and 3) CBS relay power. 1) is obvious while 2) and 3) can be understood by the fact that the performance of an AF relay system is upper bounded by the performance of both phases. We stress that the optimal cooperation scheme between the primary system and CRSs is unknown and the target rate $r_0$ may well below the maximum achievable rate.

## C. Working Modes

Cooperation is particularly essential if the primary system is in an outage. However, when the primary system is in a good state, whether cooperation is needed depends on the expected return and the complexity it involves. In the following, we outline several possible working modes.

Mode I: When the primary system is in an outage, PBS invites CBSs to cooperate. The necessary CSI such as $|\mathbf{h}_0^{p\dagger}\mathbf{f}|$, $\mathbf{g}_j$ and $\mathbf{h}_{j0}$ can either be fed back to or estimated at $\mathsf{CBS}_j$ with the help from the PBS and the PU. The CBSs will obtain the solution to (7) and make it known to the relevant terminals.[3]

Mode II: When the primary system is in a good channel state to support its rate, then

(a) Mode II.1—PBS does not need help from CBSs and gives no access of spectrum to CBSs.

(b) Mode II.2—PBS allows CBSs to access the spectrum under the condition that the interference caused by the CBSs is below some threshold $I_p$ or "interference temperature". The PBS will assist the CBSs to obtain the knowledge $\{\mathbf{h}_{j0}\}$ to ensure acceptable interference. In this case, the CBSs do not relay the PBS's signal but simply transmit their own messages with interference control to the PU. To be specific, the transmission schemes at CBSs are optimized by solving

$$\begin{aligned} &\min_{\{\mathbf{w}_m\}} \sum_{m=1}^{M} \|\mathbf{w}_m\|^2 \\ &\text{s.t.} \quad \left\{ \begin{aligned} \frac{|\mathbf{h}_{mm}^\dagger \mathbf{w}_m|^2}{\sum_{\substack{j=1\\ j\neq m}}^{M} |\mathbf{h}_{jm}^\dagger \mathbf{w}_j|^2 + N_m} &\geq \gamma_m, \\ |\mathbf{h}_{m0}^\dagger \mathbf{w}_m|^2 &\leq I_p, \forall m, \end{aligned} \right. \end{aligned} \tag{8}$$

or similarly, e.g., to maximize the CU's rate. Note that (8) has been well studied in the multiuser MIMO literature [16] and the solution is therefore omitted.

When the PU channel is in a good state, Working Mode II.1 is preferred by the PU but it is considered being selfish from the CUs' viewpoint. In contrast, the CU will prefer Mode II.2 which does not provide immediate benefit to the PU in the cooperation. Therefore, a practical scenario would be to regularly switch between the two modes to give incentives to both the PU and CU for initiating cooperation.

[3] As long as the channel remains static, the optimization is valid and the information exchange between terminals for implementing the solution should be acceptable. Re-optimization is only required if the channel changes drastically.

## III. Special Structure of $\{\mathbf{A}_m\}$ and Problem Reformulation

The problem (7) is a convex optimization problem because the objective function is convex and the constraints can also be made convex by setting the imaginary parts of $\mathbf{h}_{m0}^{\dagger}\mathbf{A}_m\mathbf{g}_m$ and $\mathbf{h}_{mm}^{\dagger}\mathbf{w}_m$ to zero [17] without loss of optimality. Therefore, (7) can be solved by the standard interior-point algorithm. However, there are convincing reasons to further study the problem to i) understand the structure that the optimal solution possesses; ii) derive more efficient algorithms than the interior-point algorithm for convex problems and iii) develop distributed implementation. The rest of this paper will be devoted to above mentioned objectives. In this section, we provide deeper understanding of the optimal solution so that physical insights can be gained and more efficient algorithms can be developed.

### A. Optimal Structure of $\{\mathbf{A}_m\}$

**Theorem 1** *The optimal* $\mathbf{A}_m$ *has the structure of*

$$\mathbf{A}_m = \mathbf{H}_m\mathbf{a}_m\mathbf{g}_m^{\dagger}, \ \forall m, \tag{9}$$

*where we have defined the composite channel matrix* $\mathbf{H}_j \triangleq [\mathbf{h}_{j0}\ \mathbf{h}_{j1} \cdots \mathbf{h}_{jM}]$ *from* $\mathsf{CBS}_j$ *to all the PU and CUs, and* $\mathbf{a}_m$ *is a complex parameter vector.*

*Proof:* See Appendix I. □

### B. Physical Insights and A Simplified Formulation

Theorem 1 states that the optimal structure of $\mathbf{A}_m$ can be divided into two components: $\mathbf{g}_m^{\dagger}$ and $\mathbf{H}_m\mathbf{a}_m$. This is quite intuitive as there is only one PU message stream and the optimal $\mathbf{A}_m$ is of rank one. It is observed that each CBS first maximizes the received SNR of the PU signal using MRC, $\mathbf{g}_j^{\dagger}$, during phase I and then relays the noisy signal using the transmit beamforming vector $\mathbf{H}_j\mathbf{a}_j$ during phase II.

This means that during phase II, the entire system resembles an interference multiple-input single-output (MISO) channel where each CBS transmits its own message as well as a noisy version of the common PU message using transmit and relay beamforming, respectively. For $\mathsf{CBS}_j$, the optimal transmit and relay beamforming vectors are both parameterized as $\mathbf{H}_j\mathbf{a}_j$ [18] where $\mathbf{a}_j$ is the parameter vector to be designed. Defining $\mathbf{v}_j \triangleq \mathbf{H}_j\mathbf{a}_j$, the received PU signal (without noise) via $\mathsf{CBS}_j$ is expressed as

$$\hat{s}_0 = \mathbf{h}_{j0}^{\dagger}\left[\mathbf{v}_j\mathbf{g}_j^{\dagger}(\mathbf{g}_j s_0 + \mathbf{n}_{sj})\right] = \|\mathbf{g}_j\|^2\mathbf{h}_{j0}^{\dagger}\mathbf{v}_j s_0 + \mathbf{h}_{j0}^{\dagger}\mathbf{v}_j\mathbf{g}_j^{\dagger}\mathbf{n}_{sj}. \tag{10}$$

The overall relaying operation is illustrated in Fig. 2. Note that all CBSs should form a collaborative-relaying beam to ensure coherent reception at the PU, i.e., $\mathbf{h}_{j0}^{\dagger}\mathbf{v}_j$ $\forall j$ are co-phased. When CBSs act purely as relays without serving their users, the structures of $\{\mathbf{A}_m\}$ coincide with the results in [19].

Each $\mathbf{A}_j$ originally is a general $N \times N$ matrix but (9) indicates that it is a rank-1 matrix and can be represented by a relaying vector $\mathbf{v}_j$ with dimension $N$. As a result, substituting (9) into (7), we get

$$\min_{\{\mathbf{w}_j,\mathbf{v}_j\}} \quad \sum_{j=1}^{M}\|\mathbf{w}_j\|^2 + \sum_{j=1}^{M}(P_0\|\mathbf{g}_j\|^2 + N_{sj})\|\mathbf{g}_j\|^2\|\mathbf{v}_j\|^2 \tag{11a}$$

$$\text{s.t.} \quad \begin{cases} \dfrac{|\sum_{j=1}^{M}\mathbf{h}_{j0}^{\dagger}\mathbf{v}_j\|\mathbf{g}_j\|^2|^2}{\sum_{j=1}^{M}|\mathbf{h}_{j0}^{\dagger}\mathbf{w}_j|^2 + \sum_{j=1}^{M}N_{sj}|\mathbf{h}_{j0}^{\dagger}\mathbf{v}_j|^2\|\mathbf{g}_j^{\dagger}\|^2 + N_2^p} \geq \gamma_0^{'}, \\ \dfrac{|\mathbf{h}_{mm}^{\dagger}\mathbf{w}_m|^2}{\sum_{\substack{j=1\\ j\neq m}}^{M}|\mathbf{h}_{jm}^{\dagger}\mathbf{w}_j|^2 + \sum_{j=1}^{M}(P_0\|\mathbf{g}_j\|^2 + N_{sj})\|\mathbf{g}_j\|^2|\mathbf{h}_{jm}^{\dagger}\mathbf{v}_j|^2 + N_m} \geq \gamma_m, \forall m. \end{cases} \tag{11b}$$

For notational convenience, we define the following vectors and matrices:

$$\begin{cases} \mathbf{v} \triangleq \mathsf{vec}(\mathbf{v}_1,\ldots,\mathbf{v}_M), \\ \mathbf{h}_0 \triangleq [\mathbf{h}_{10}\|\mathbf{g}_1\|^2;\cdots;\mathbf{h}_{M0}\|\mathbf{g}_M\|^2], \text{where ``;'' denotes vertical concatenation of vectors,} \\ \bar{\mathbf{H}}_0 \triangleq \mathsf{BlkDiag}\left(\mathsf{vec}\left(\mathbf{h}_{10}^{\dagger}\sqrt{N_{s1}}\|\mathbf{g}_1\|,\ldots,\mathbf{h}_{M0}^{\dagger}\sqrt{N_{sM}}\|\mathbf{g}_M\|\right)\right), \\ \bar{\mathbf{H}}_m \triangleq \mathsf{BlkDiag}\left(\mathsf{vec}\left(\mathbf{h}_{1m}^{\dagger}\|\mathbf{g}_1\|\sqrt{(P_0\|\mathbf{g}_1\|^2 + N_{s1})},\ldots,\mathbf{h}_{Mm}^{\dagger}\|\mathbf{g}_M\|\sqrt{(P_0\|\mathbf{g}_M\|^2 + N_{sM})}\right)\right), \forall m, \\ \mathbf{D}_v \text{ is a diagonal matrix with its } (n,n)\text{-th entry being } (P_0\|\mathbf{g}_j\|^2 + N_{sj})\|\mathbf{g}_j\|^2, \\ \qquad\qquad \text{for } n = (j-1)M+1,\ldots,jM. \end{cases} \tag{12}$$

Then, (11) can be rewritten as

$$\min_{\{\mathbf{w}_j,\mathbf{v}\}} \sum_{j=1}^{M}\|\mathbf{w}_j\|^2 + \mathbf{v}^{\dagger}\mathbf{D}_v\mathbf{v} \;\; \text{s.t.} \;\; \begin{cases} \dfrac{|\mathbf{h}_0^{\dagger}\mathbf{v}|^2}{\sum_{j=1}^{M}|\mathbf{h}_{j0}^{\dagger}\mathbf{w}_j|^2 + \|\bar{\mathbf{H}}_0\mathbf{v}\|^2 + N_2^p} \geq \gamma_0^{'}, \\ \dfrac{|\mathbf{h}_{mm}^{\dagger}\mathbf{w}_m|^2}{\sum_{\substack{j=1\\ j\neq m}}^{M}|\mathbf{h}_{jm}^{\dagger}\mathbf{w}_j|^2 + \|\bar{\mathbf{H}}_m\mathbf{v}\|^2 + N_m} \geq \gamma_m, \forall m. \end{cases} \tag{13}$$

Like (7) before, (13) can be equivalently converted into a convex problem.

# IV. Efficient Algorithms and Implementations

In this section, we propose two algorithms to solve the dual problem of (13), which can be derived from its Karush-Kuhn-Tucker (KKT) conditions and is expressed as

$$\begin{aligned}
&\max_{\lambda_0,\{\lambda_m\}\geq 0} \quad \lambda_0 N_2^p + \sum_{m=1}^{M} \lambda_m N_m \\
&\text{s.t.} \quad \begin{cases} \mathbf{I} + \lambda_0 \mathbf{h}_{m0}\mathbf{h}_{m0}^{\dagger} + \sum\limits_{\substack{n=1\\ n\neq m}}^{M} \lambda_n \mathbf{h}_{mn}\mathbf{h}_{mn}^{\dagger} \succeq \dfrac{\lambda_m}{\gamma_m}\mathbf{h}_{mm}\mathbf{h}_{mm}^{\dagger}, \forall m, \\ \mathbf{D}_v + \sum\limits_{m=1}^{M} \lambda_m \bar{\mathbf{H}}_m^{\dagger}\bar{\mathbf{H}}_m + \lambda_0 \bar{\mathbf{H}}_0^{\dagger}\bar{\mathbf{H}}_0 \succeq \dfrac{\lambda_0}{\gamma_0'}\mathbf{h}_0\mathbf{h}_0^{\dagger}. \end{cases}
\end{aligned} \tag{14}$$

where $\lambda_0, \{\lambda_m\}$ are dual variables and their meanings will be explained later. The procedures to derive it can be found in the textbook [20] and previous work [17]. We will then discuss in Section IV-C how to recover the solution of the original problem after the dual problem is solved.

## A. Algorithm 1 Based on (14) and Fixed-Point Iteration

The first proposed algorithm is directly based on the dual problem formulation (14). To proceed, we first derive an equivalent form of the constraints by rewriting the constraint in (14) as

$$\mathbf{I} - \left(\mathbf{I} + \lambda_0 \mathbf{h}_{m0}\mathbf{h}_{m0}^{\dagger} + \sum_{\substack{n=1\\ n\neq m}}^{M} \lambda_n \mathbf{h}_{mn}\mathbf{h}_{mn}^{\dagger}\right)^{-\frac{1}{2}} \frac{\lambda_m}{\gamma_m}\mathbf{h}_{mm}\mathbf{h}_{mm}^{\dagger} \left(\mathbf{I} + \lambda_0 \mathbf{h}_{m0}\mathbf{h}_{m0}^{\dagger} + \sum_{\substack{n=1\\ n\neq m}}^{M} \lambda_n \mathbf{h}_{mn}\mathbf{h}_{mn}^{\dagger}\right)^{-\frac{1}{2}} \succeq \mathbf{0}, \tag{15}$$

and further

$$\lambda_m \leq \frac{\gamma_m}{\mathbf{h}_{mm}^{\dagger}\left(\mathbf{I} + \lambda_0 \mathbf{h}_{m0}\mathbf{h}_{m0}^{\dagger} + \sum_{\substack{n=1\\ n\neq m}}^{M} \lambda_n \mathbf{h}_{mn}\mathbf{h}_{mn}^{\dagger}\right)^{-1}\mathbf{h}_{mm}}. \tag{16}$$

Since the objective function increases with $\lambda_m$, at the optimum, the equality must hold; similarly for $\lambda_0$.

**Algorithm 1: Efficient fixed-point iteration to solve (14).**

Step 1) Initialize $\boldsymbol{\lambda} \triangleq [\lambda_0, \lambda_1, \ldots, \lambda_M]^T$ arbitrarily.

Step 2) Update

$$\lambda_m(\boldsymbol{\lambda}) = \frac{\gamma_m}{\mathbf{h}_{mm}^{\dagger}\left(\mathbf{I} + \lambda_0 \mathbf{h}_{m0}\mathbf{h}_{m0}^{\dagger} + \sum_{\substack{n=1\\ n\neq m}}^{M} \lambda_n \mathbf{h}_{mn}\mathbf{h}_{mn}^{\dagger}\right)^{-1}\mathbf{h}_{mm}}, \ \forall m \geq 1. \tag{17}$$

Step 3) Update

$$\lambda_0(\boldsymbol{\lambda}) = \frac{\gamma_0^{'}}{\mathbf{h}_0^{\dagger}\left(\mathbf{D}_v + \sum_{m=1}^{M}\lambda_m\bar{\mathbf{H}}_m^{\dagger}\bar{\mathbf{H}}_m + \lambda_0\bar{\mathbf{H}}_0^{\dagger}\bar{\mathbf{H}}_0\right)^{-1}\mathbf{h}_0}. \tag{18}$$

Step 4) Go back to Step 2) until convergence.

**Theorem 2** *Algorithm 1 finds the optimal solution to (14).*

*Proof:* It is easy to check that the mappings in (17) and (18) satisfy

- Positivity: $\lambda_m(\boldsymbol{\lambda}) \geq \mathbf{0}, \lambda_0(\boldsymbol{\lambda}) \geq \mathbf{0}$;
- Monotonicity: If $\boldsymbol{\lambda} \geq \boldsymbol{\lambda}^{'}$, then $\lambda_m(\boldsymbol{\lambda}) \geq \lambda_m(\boldsymbol{\lambda}^{'}), \lambda_0(\boldsymbol{\lambda}) \geq \lambda_0(\boldsymbol{\lambda}^{'})$;
- Scalability: For all $\alpha > 1$, $\alpha\lambda_m(\boldsymbol{\lambda}) > \lambda_m(\alpha\boldsymbol{\lambda}), \alpha\lambda_0(\boldsymbol{\lambda}) > \lambda_0(\alpha\boldsymbol{\lambda})$.

Thus, they are standard interference functions and the optimality follows directly the results in [21]. □

### B. Algorithm 2 Based on the Virtual Uplink and Matrix Iteration

In Phase II, as the channel from the CBSs to all CUs and PU resembles a multicell downlink [22], we propose an even more efficient algorithm by exploiting the duality between the original downlink and a virtual uplink. In this case, however, we assume that there is one CBS or an additional central unit who coordinates the optimization, i.e., a centralized approach. Based on (14), we construct an equivalent power optimization problem in a virtual uplink by introducing the auxiliary vectors $\{\tilde{\mathbf{w}}_m, \tilde{\mathbf{v}}\}$ as

$$\begin{aligned} &\max_{\boldsymbol{\lambda}\geq\mathbf{0}} \ \mathbf{n}^T\boldsymbol{\lambda} \\ &\text{s.t.} \begin{cases} \max\limits_{\tilde{\mathbf{w}}_m} \dfrac{\lambda_m|\mathbf{h}_{mm}^{\dagger}\tilde{\mathbf{w}}_m|^2}{\lambda_0|\mathbf{h}_{m0}^{\dagger}\tilde{\mathbf{w}}_m|^2 + \sum_{\substack{n=1\\n\neq m}}^{M}\lambda_n|\mathbf{h}_{mn}^{\dagger}\tilde{\mathbf{w}}_m|^2 + \|\tilde{\mathbf{w}}_m^{\dagger}\|^2} \leq \gamma_m, \forall m, \\ \max\limits_{\tilde{\mathbf{v}}} \dfrac{\lambda_0|\mathbf{h}_0^{\dagger}\tilde{\mathbf{v}}|^2}{\tilde{\mathbf{v}}^{\dagger}\mathbf{D}_v\tilde{\mathbf{v}} + \sum_{m=1}^{M}\lambda_m\|\bar{\mathbf{H}}_m\tilde{\mathbf{v}}\|^2 + \lambda_0\|\bar{\mathbf{H}}_0\tilde{\mathbf{v}}\|^2} \leq \gamma_0^{'}, \end{cases} \end{aligned} \tag{19}$$

where $\mathbf{n} \triangleq [N_2^p \ N_1 \cdots N_m]^T$. Also, the vectors $\{\tilde{\mathbf{w}}_m\}$ and $\tilde{\mathbf{v}}$ can be interpreted as the receive beamforming and relay beamforming vectors in the virtual uplink, respectively.

Since all the constraints in (19) should be satisfied with equality at the optimum, all the inequality signs can be reversed and the maximization can be replaced by minimization of the objective function. As a consequence, (19) can be equivalently rewritten as

$$
\begin{aligned}
\min_{\boldsymbol{\lambda}\geq\mathbf{0},\{\tilde{\mathbf{w}}_m\},\tilde{\mathbf{v}}} \quad & \mathbf{n}^T\boldsymbol{\lambda} \\
\text{s.t.} \quad & \begin{cases} \dfrac{\lambda_m|\mathbf{h}_{mm}^{\dagger}\tilde{\mathbf{w}}_m|^2}{\lambda_0|\mathbf{h}_{m0}^{\dagger}\tilde{\mathbf{w}}_m|^2+\sum_{n=1,n\neq m}^{M}\lambda_n|\mathbf{h}_{mn}^{\dagger}\tilde{\mathbf{w}}_m|^2+\|\tilde{\mathbf{w}}_m^{\dagger}\|^2}\geq\gamma_m,\forall m, \\ \dfrac{\lambda_0|\mathbf{h}_0^{\dagger}\tilde{\mathbf{v}}|^2}{\tilde{\mathbf{v}}^{\dagger}\mathbf{D}_v\tilde{\mathbf{v}}+\sum_{m=1}^{M}\lambda_m\|\bar{\mathbf{H}}_m\tilde{\mathbf{v}}\|^2+\lambda_0\|\bar{\mathbf{H}}_0\tilde{\mathbf{v}}\|^2}\geq\gamma_0^{'}. \end{cases}
\end{aligned} \tag{20}
$$

The "max" operations in the constraints of (19) are removed as they are automatically enforced in (20). The equivalence of (19) and (20) can be verified by showing that a feasible solution of one is also feasible for the other. Now, we have a power minimization problem (20) in the virtual uplink where $\lambda_m$ can be interpreted as the uplink transmit power for $\mathsf{CU}_m$, and $\lambda_0$ is the total relaying power over all the CBSs. Intriguingly, instead of minimizing the total CBS transmit power in the downlink, we solve it by minimizing the total power of relaying and the CUs in the virtual uplink. Also, $\tilde{\mathbf{w}}_m$ becomes the receive beamforming vector at $\mathsf{CBS}_m$ to recover the message from $\mathsf{CU}_m$ while $\tilde{\mathbf{v}}$ is regarded as the receive relaying vector.

**Corollary 1** *The same SINR region $\{\gamma_0^{'},\gamma_1,\ldots,\gamma_M\}$ can be achieved in the downlink and the virtual uplink described by (20), using the same set of normalized transmit/receive beamforming vectors and relaying vectors. Mathematically, that is, in (11) and (20), we have $\mathbf{v}\propto\tilde{\mathbf{v}}$ and $\mathbf{w}_m\propto\tilde{\mathbf{w}}_m,\forall m$.*

With Corollary 1, it suffices to solve (20). Before doing so, we define:

$$
\mathbf{F}\in\mathbb{R}^{(M+1)\times(M+1)}:\begin{cases} \mathbf{F}_{1,1}=\|\bar{\mathbf{H}}_0\tilde{\mathbf{v}}\|^2, \\ \mathbf{F}_{n,n}=0,\ \text{for } n=2,\ldots,M+1, \\ \mathbf{F}_{n+1,1}=|\mathbf{h}_{n0}^{\dagger}\tilde{\mathbf{w}}_n|^2,\ \text{for } n=1,\ldots,M, \\ \mathbf{F}_{1,n+1}=\|\bar{\mathbf{H}}_m\tilde{\mathbf{v}}\|^2,\ \text{for } n=1,\ldots,M, \\ \mathbf{F}_{m+1,n+1}=|\mathbf{h}_{nm}^{\dagger}\tilde{\mathbf{w}}_m|^2,\ \text{for } m,n=1,\ldots,M,\ \text{and } m\neq n, \end{cases} \tag{21}
$$

$$
\mathbf{D}\in\mathbb{R}^{(M+1)\times(M+1)}:\begin{cases} \mathbf{D}_{1,1}=\dfrac{\gamma_0^{'}}{|\mathbf{h}_0^{\dagger}\tilde{\mathbf{v}}|^2}, \\ \mathbf{D}_{n,n}=\dfrac{\gamma_n}{|\mathbf{h}_{nn}\tilde{\mathbf{w}}_n|^2},\ \text{for } n=2,\ldots,M+1, \\ \mathbf{D}_{m,n}=0,\ \text{for } m\neq n, \end{cases} \tag{22}
$$

$$\boldsymbol{\sigma} \in \mathbb{R}^{(M+1)\times 1}: \begin{cases} \boldsymbol{\sigma}_1 = \tilde{\mathbf{v}}^\dagger \mathbf{D}_v \tilde{\mathbf{v}}, \\ \boldsymbol{\sigma}_n = 1, \text{ for } n = 2, \ldots, M+1. \end{cases} \tag{23}$$

Then, (20) can be written as

$$\min_{\boldsymbol{\lambda} \geq \mathbf{0}, \{\tilde{\mathbf{w}}_m\}, \tilde{\mathbf{v}}} \mathbf{n}^T \boldsymbol{\lambda} \text{ s.t. } \boldsymbol{\lambda} \geq \mathbf{D}\mathbf{F}\boldsymbol{\lambda} + \mathbf{D}\boldsymbol{\sigma}. \tag{24}$$

Notice that at the optimum, the constraint in (24) becomes an equality constraint.

**Algorithm 2: Efficient matrix iteration to solve (24).**

Step 1) Find a feasible solution $(\boldsymbol{\lambda}, \{\tilde{\mathbf{w}}_m\}, \tilde{\mathbf{v}})$ to satisfy the SINR constraints in (24).

Step 2) Update

$$\boldsymbol{\lambda} = (\mathbf{I} - \mathbf{D}\mathbf{F})^{-1}\mathbf{D}\boldsymbol{\sigma}. \tag{25}$$

Step 3) Update

$$\tilde{\mathbf{w}}_m = \left( \mathbf{I} + \lambda_0 \mathbf{h}_{m0}\mathbf{h}_{m0}^\dagger + \sum_{\substack{n=1 \\ n\neq m}}^{M} \lambda_n \mathbf{h}_{mn}\mathbf{h}_{mn}^\dagger \right)^{-1} \mathbf{h}_{mm}, \quad \text{and} \quad \tilde{\mathbf{w}}_m = \frac{\tilde{\mathbf{w}}_m}{\|\tilde{\mathbf{w}}_m\|}, \forall m. \tag{26}$$

Step 4) Update

$$\tilde{\mathbf{v}} = \left( \mathbf{D}_v + \sum_{m=1}^{M} \lambda_m \bar{\mathbf{H}}_m^\dagger \bar{\mathbf{H}}_m + \lambda_0 \bar{\mathbf{H}}_0^\dagger \bar{\mathbf{H}}_0 \right)^{-1} \mathbf{h}_0, \quad \text{and} \quad \tilde{\mathbf{v}} = \frac{\tilde{\mathbf{v}}}{\|\tilde{\mathbf{v}}\|}. \tag{27}$$

Step 5) Go back to Step 2) until convergence.

The optimality of Algorithm 2 has been established in [23] provided a feasible initial solution is given. To obtain a feasible solution in Step 1, we study the following SINR balancing problem:

$$\begin{aligned} \max_{C, \boldsymbol{\lambda} \geq \mathbf{0}, \{\tilde{\mathbf{w}}_m\}, \tilde{\mathbf{v}}} \quad & C \\ \text{s.t.} \quad & \begin{cases} \dfrac{\boldsymbol{\lambda}}{C} = \mathbf{D}\mathbf{F}\boldsymbol{\lambda} + \mathbf{D}\boldsymbol{\sigma}, \\ \mathbf{n}^T \boldsymbol{\lambda} \leq P_T, \end{cases} \end{aligned} \tag{28}$$

where $P_T$ is the total power constraint. (20) is feasible if and only if the optimal $C$ satisfies $C \geq 1$. It can be checked that the first matrix equation constraint in (28) has a similar mathematical structure as [23, (34)]. Thus, the algorithm based on an eigen-system proposed in [23, Section IV-C] can be applied to solve (28), which results in the first step of Algorithm 2. If a feasible solution is found, the rest of Algorithm 2 solves the power minimization problem (24) via matrix iteration following [23, Section III-C].

## C. Conversion to the Original Downlink Solutions

After obtaining the optimal solution $\boldsymbol{\lambda}$ to the dual problem (14), we can find the unit-norm beamforming vectors using (26) and (27). Having identified $\{\tilde{\mathbf{w}}_m\}$ and $\tilde{\mathbf{v}}$, we can decide on the downlink power based on the fact that at the optimum, all constraints in (7) are met with equalities; therefore we have $M+1$ independent linear equations sufficient to solve for the $(M+1)\times 1$ power vector to get

$$\mathbf{p} \triangleq [p_0; p_1; \cdots; p_M] = (\mathbf{I} - \mathbf{D}\mathbf{F}^T)^{-1}\mathbf{D}\boldsymbol{\sigma}. \tag{29}$$

As a result, the downlink solution is given by

$$\begin{cases} \mathbf{v} = \sqrt{p_0}\tilde{\mathbf{v}}, \\ \mathbf{w}_m = \sqrt{p_m}\tilde{\mathbf{w}}_m, \forall m. \end{cases} \tag{30}$$

A straightforward implementation of the solution is to make the matrix $(\mathbf{I} - \mathbf{D}\mathbf{F}^T)^{-1}\mathbf{D}$ known to all the CBSs. Alternatively, however, if the beamforming directions are known, then it is actually possible to obtain the optimal downlink power locally in a distributed manner which avoids the matrix inversion.[4]

## D. Comparisons Between the Two Algorithms

### D.1 Convergence Rate Analysis and Complexity Comparison

Comparing Algorithm 1 and Algorithm 2, the only difference lies in the way the power is being updated. In Algorithm 1, the power optimizations (17) and (18) are based on the fixed point iteration while in Algorithm 2, the power is optimized based on the matrix equation (25). Assuming that the beamforming vectors are fixed, the power iteration in Algorithm 1 can be rewritten in a matrix form as

$$\boldsymbol{\lambda}^{(n+1)} = \mathbf{D}\mathbf{F}\boldsymbol{\lambda}^{(n)} + \mathbf{D}\boldsymbol{\sigma}, \tag{31}$$

where $\boldsymbol{\lambda}^{(n)}$ denotes the solution at the $n$-th iteration.

In [25], it has been analyzed that such iteration satisfies

$$\sup_{x>0} \frac{\|\boldsymbol{\lambda}^{(n+1)} - \boldsymbol{\lambda}^*\|_x}{\|\boldsymbol{\lambda}^{(n)} - \boldsymbol{\lambda}^*\|_x} \geq \rho(\mathbf{D}\mathbf{F}), \tag{32}$$

[4] To do so, each user feeds back information about its received SINR given a power solution, and then adjusts its transmit power accordingly to meet its target SINR. This distributed power control is known to achieve the optimum [24].

where $\boldsymbol{\lambda}^*$ is the optimal solution and $\rho(\cdot)$ returns the spectral radius of an input matrix. The result (32) indicates that the power update in Algorithm 1 has linear convergence, which is true even the beamforming vectors are adapted at each iteration. On the other hand, in Algorithm 2, the power update directly optimizes the objective function to exactly satisfy all the SINR constraints with equality, i.e.,

$$\boldsymbol{\lambda} = (\mathbf{I} - \mathbf{D}\mathbf{F})^{-1}\mathbf{D}\boldsymbol{\sigma}. \tag{33}$$

Hence, it is expected that starting with the same initial point, at each iteration, the power vector generated by Algorithm 2 is element-wisely less than that generated by Algorithm 1, which has been formally proved in [25] and it is further shown that the power sequence generated by Algorithm 2 satisfies

$$\lim_{n\to\infty} \sup_{x>0} \frac{\|\boldsymbol{\lambda}^{(n+1)} - \boldsymbol{\lambda}^*\|_x}{\|\boldsymbol{\lambda}^{(n)} - \boldsymbol{\lambda}^*\|_x} = 0, \tag{34}$$

which indicates that Algorithm 2 has a superlinear convergence rate. This comparison helps explain why Algorithm 2 converges faster than Algorithm 1, and will be verified by simulation results in Section VI.

For complexity comparison, note that (17)–(18) in Algorithm 1 and (26)–(27) in Algorithm 2 involve similar matrix inversion operations, so their complexities are comparable. Compared with Algorithm 1, Algorithm 2 needs additional Step 1) to find the feasible solution via eigenvalue decomposition and Step 2) to find the optimal $\lambda$ via matrix inversion (25). These steps have the overall complexity $\mathcal{O}(M^3)$.

### D.2 Distributed Implementation

For implementation, the power update (33) in Algorithm 2 requires both global CSI and centralized processing. On the contrary, from (31) in Algorithm 1, each CU's power can be updated locally assuming all others are fixed, which gives rise to a distributed implementation. This is described as follows.

The optimization of $\lambda_m$ in (17) of Algorithm 1 requires only the CSI from $\mathsf{CBS}_m$ to all the CUs and the PU, i.e., $\mathbf{H}_m = [\mathbf{h}_{m0}\ \mathbf{h}_{m1} \cdots \mathbf{h}_{mM}]$ defined in Appendix I, and therefore can be implemented at $\mathsf{CBS}_m$ locally. To obtain $\boldsymbol{\lambda}$ other than $\lambda_m$ ($\lambda_0$ will be discussed next), it can be obtained by a small amount of information exchange among all the CBSs (see [15] for details) or feedback from the CUs [21]. After the optimal $\boldsymbol{\lambda}$ is known, the optimization of $\mathbf{w}_m$ can be readily achieved using (26) with local CSI.

The optimization of $\lambda_0$ in (18) of Algorithm 1 is less straightforward. At first, it may seem that the update of $\lambda_0$ involves global CSI, but in fact the matrices $\bar{\mathbf{H}}_m^\dagger \bar{\mathbf{H}}_m$ and $\bar{\mathbf{H}}_0^\dagger \bar{\mathbf{H}}_0$ are both block diagonal, and therefore the right-hand-side can be decoupled. After some manipulation, we can get

$$\mathbf{h}_0^\dagger \left( \mathbf{D}_v + \sum_{m=1}^{M} \lambda_m \bar{\mathbf{H}}_m^\dagger \bar{\mathbf{H}}_m + \lambda_0 \bar{\mathbf{H}}_0^\dagger \bar{\mathbf{H}}_0 \right)^{-1} \mathbf{h}_0 = \sum_{j=1}^{M} \alpha_j, \tag{35}$$

where

$$\alpha_j \triangleq \|\mathbf{g}_j\|^2 \mathbf{h}_{j0}^\dagger \left( \lambda_0 N_{sj} \mathbf{h}_{j0} \mathbf{h}_{j0}^\dagger + (P_0 \|\mathbf{g}_j\|^2 + N_{sj}) \mathbf{I} + \sum_{m=1}^{M} \lambda_m (P_0 \|\mathbf{g}_j\|^2 + N_{sj}) \mathbf{h}_{jm} \mathbf{h}_{jm}^\dagger \right)^{-1} \mathbf{h}_{j0}. \tag{36}$$

Note that the SINR is the summation of contributions from all the collaborative CBSs. For $\mathsf{CBS}_j$, $\alpha_j$ can be evaluated using local CSI and $\boldsymbol{\lambda}$. Thus, the update of $\lambda_0$ can be done in a distributed manner by

$$\lambda_0 = \frac{\gamma_0^{'}}{\sum_{j=1}^{M} \alpha_j}, \tag{37}$$

which requires that the CBSs exchange information of $\{\alpha_j\}$ with each other.

Once the optimal $\boldsymbol{\lambda}$ is found, $\mathsf{CBS}_j$ calculates the relay-beamforming vector by

$$\mathbf{v}_j = \|\mathbf{g}_j\|^2 \left( \lambda_0 N_{sj} \mathbf{h}_{j0} \mathbf{h}_{j0}^\dagger + (P_0 \|\mathbf{g}_j\|^2 + N_{sj}) \mathbf{I} + \sum_{m=1}^{M} \lambda_m (P_0 \|\mathbf{g}_j\|^2 + N_{sj}) \mathbf{h}_{jm} \mathbf{h}_{jm}^\dagger \right)^{-1} \mathbf{h}_{j0}, \tag{38}$$

which again only relies on local CSI and can be implemented in a distributed manner.

The structure of $\mathbf{v}_j$ is intuitive. If the CBSs only act as relays without serving their CUs, the optimal collaborative-relaying strategy is to let $\mathbf{v}_j$ match the forward channel $\{\mathbf{h}_{j0}\}$. From (38), we have

$$\mathbf{v}_j \propto \left( \mathbf{I} + \sum_{m=1}^{M} \lambda_m \mathbf{h}_{jm} \mathbf{h}_{jm}^\dagger \right)^{-1} \mathbf{h}_{j0}. \tag{39}$$

While the CBSs also send messages to their CUs which causes interference to the PU, $\mathbf{v}_j$ plays a role to form a collaborative signaling beam to compensate the effects of interference.

To summarize, Algorithm 1 is desirable for distributed implementation when a central controller is absent to gather global CSI and coordinate the required optimization. With only local CSI at each CBS and moderate amount of parameter exchange, the optimal transmit and relaying beamforming can be realized in a distributed manner. Next, we analyze the signalling requirement of the algorithms.

### D.3 Discussion on Parameter Exchange

The distributed strategy depends largely on the required parameter exchange between the CBSs in two steps. First, Algorithm 1 is used to solve the dual problem and at each iteration, the vectors $[\lambda_1, \dots, \lambda_M]$ and $\boldsymbol{\alpha} = [\alpha_1, \dots, \alpha_M]$ need to be shared among the CBSs. The total number of exchanged parameters is $2N_I M$ where $N_I$ denotes the total number of iterations needed for Algorithm 1 to converge. Then, for the uplink-downlink conversion, the matrix $(\mathbf{I} - \mathbf{D}\mathbf{F}^T)^{-1}\mathbf{D}$ needs to be made available to all the CBSs, requiring $(M+1)^2$ positive scalars to be shared. The total number of parameter exchange is therefore $2N_I M + (M+1)^2$, regardless of the number of transmit antennas, $N$. As a comparison, full CSI exchange, which would make possible a centralized optimization (e.g., Algorithm 2), will require $2NM^2$ scalars to be shared where the factor of 2 is due to the complex nature of the parameters. If the same quantization scheme is used, the ratio of the number of exchanged bits in Algorithm 1 over full CSI exchange is

$$\eta = \frac{2N_I M + (M+1)^2}{2NM^2} \approx \frac{N_I + \frac{1}{2}}{NM}. \tag{40}$$

When the number of CBSs, $M$, is large, a great reduction in backhaul signaling is anticipated.

## V. Two Zero-forcing Solutions

This section aims to develop two low-complexity *closed-form* zero-forcing solutions to (7).

### A. Transmit Zero-forcing Beamformer

For coordinated beamforming of the CBSs, we write $\mathbf{w}_m = \sqrt{p_m}\bar{\mathbf{w}}_m$, where $\|\bar{\mathbf{w}}_m\| = 1$. For zero-forcing, $\bar{\mathbf{w}}_m$ should be chosen such that there is no interference from $\mathsf{CBS}_m$ to the PU and all other CUs, i.e.,

$$\bar{\mathbf{w}}_m^\dagger \mathbf{H}_m = \mathbf{0}, \forall m, \tag{41}$$

where $\mathbf{H}_m = [\mathbf{h}_{m0}\ \mathbf{h}_{m1}\ \cdots\ \mathbf{h}_{mm-1}\ \mathbf{h}_{mm+1}\ \cdots\ \mathbf{h}_{mM}]$ has been defined in Appendix I as the composite forward channel from $\mathsf{CBS}_j$ to all the PU and CUs. One possible solution for the transmit zero-forcing beamforming is expressed as

$$\bar{\mathbf{w}}_m = \frac{\left(\mathbf{I} - \mathbf{H}_m(\mathbf{H}_m^\dagger\mathbf{H}_m)^{-1}\mathbf{H}_m^\dagger\right)\mathbf{h}_{mm}}{\left\|\left(\mathbf{I} - \mathbf{H}_m(\mathbf{H}_m^\dagger\mathbf{H}_m)^{-1}\mathbf{H}_m^\dagger\right)\mathbf{h}_{mm}\right\|}, \forall m. \tag{42}$$

With the above solution, the original problem becomes

$$\min_{\{p_j,\mathbf{v}_j\}} \quad \sum_{j=1}^{M} p_j + \sum_{j=1}^{M} (P_0\|\mathbf{g}_j\|^2 + N_{sj})\|\mathbf{g}_j\|^2\|\mathbf{v}_j\|^2 \tag{43a}$$

$$\text{s.t.} \quad \begin{cases} \dfrac{|\sum_{j=1}^{M} \mathbf{h}_{j0}^{\dagger}\mathbf{v}_j\|\mathbf{g}_j\|^2|^2}{\sum_{j=1}^{M} N_{sj}|\mathbf{h}_{j0}^{\dagger}\mathbf{v}_j|^2\|\mathbf{g}_j^{\dagger}\|^2 + N_2^p} \geq \gamma_0^{'}, \\ \dfrac{p_m|\mathbf{h}_{mm}^{\dagger}\bar{\mathbf{w}}_m|^2}{\sum_{j=1}^{M}(P_0\|\mathbf{g}_j\|^2 + N_{sj})\|\mathbf{g}_j\|^2|\mathbf{h}_{jm}^{\dagger}\mathbf{v}_j|^2 + N_m} \geq \gamma_m, \forall m. \end{cases} \tag{43b}$$

## B. CZF Relaying Vectors

With CZF, the design of $\mathbf{v}_j$ that ensures no interference to all the CUs should satisfy

$$\bar{\mathbf{v}}_j^{\dagger}\mathbf{H}_{cm} = \mathbf{0}, \tag{44}$$

where $\mathbf{H}_{cm} \triangleq [\mathbf{h}_{j1}, \ldots, \mathbf{h}_{jM}]$ and $\bar{\mathbf{v}}_j$ is chosen to maximize the received SINR for the PU so that

$$\bar{\mathbf{v}}_j = \left(\mathbf{I} - \mathbf{H}_{cm}(\mathbf{H}_{cm}^{\dagger}\mathbf{H}_{cm})^{-1}\mathbf{H}_{cm}^{\dagger}\right)\mathbf{h}_{j0}\|\mathbf{g}_j\|^2. \tag{45}$$

Note that (45) also ensures that all $\mathbf{h}_{j0}^{\dagger}\mathbf{v}_j$ are co-phased for coherent reception. Defining $\bar{\mathbf{v}} = \frac{[\bar{\mathbf{v}}_1;\cdots;\bar{\mathbf{v}}_M]}{\|[\bar{\mathbf{v}}_1;\cdots;\bar{\mathbf{v}}_M]\|}$ to have $\mathbf{v} = \sqrt{p_0}\bar{\mathbf{v}}$, (43) is then reduced to the following power minimization problem:

$$\begin{aligned} \min_{\{p_j,p_0\}} \quad & \sum_{j=1}^{M} p_j + p_0\bar{\mathbf{v}}^{\dagger}\mathbf{D}_v\bar{\mathbf{v}} \\ \text{s.t.} \quad & \begin{cases} \dfrac{p_0|\mathbf{h}_{m0}^{\dagger}\bar{\mathbf{v}}|^2}{p_0\|\bar{\mathbf{H}}_0\bar{\mathbf{v}}\|^2 + N_2^p} \geq \gamma_0^{'} \\ p_m\dfrac{|\mathbf{h}_{mm}^{\dagger}\bar{\mathbf{w}}_m|^2}{N_m} \geq \gamma_m, \forall m. \end{cases} \end{aligned} \tag{46}$$

From the constraints in (46), we can determine the minimum required transmit power as

$$\begin{cases} p_0 = \dfrac{\gamma_0^{'}N_2^p}{|\mathbf{h}_{m0}^{\dagger}\bar{\mathbf{v}}|^2 - \gamma_0^{'}\|\bar{\mathbf{H}}_0\bar{\mathbf{v}}\|^2} \\ p_m = \dfrac{\gamma_m N_m}{|\mathbf{h}_{mm}^{\dagger}\bar{\mathbf{w}}_m|^2}, \forall m. \end{cases} \tag{47}$$

## C. PZF Relaying Vectors

In PZF [13], $\mathbf{v}$ aims to maximize the PU's channel without controlling the interference to CUs, i.e.,

$$\bar{\mathbf{v}} = \frac{[\mathbf{h}_{10}; \cdots ; \mathbf{h}_{M0}]}{\|[\mathbf{h}_{10}; \cdots ; \mathbf{h}_{M0}]\|}. \tag{48}$$

This gives priority to the PU's transmission at the expense of the CUs. With (48), (43) becomes

$$\min_{\{p_j, p_0\}} \sum_{j=1}^{M} p_j + p_0 \bar{\mathbf{v}}^\dagger \mathbf{D}_v \bar{\mathbf{v}} \quad \text{s.t.} \quad \begin{cases} \dfrac{p_0 |\mathbf{h}_{m0}^\dagger \bar{\mathbf{v}}|^2}{p_0 \|\bar{\mathbf{H}}_0 \bar{\mathbf{v}}\|^2 + N_2^p} \geq \gamma_0^{'}, \\ \dfrac{p_m |\mathbf{h}_{mm}^\dagger \bar{\mathbf{w}}_m|^2}{p_0 \|\bar{\mathbf{H}}_m \bar{\mathbf{v}}\|^2 + N_m} \geq \gamma_m, \forall m. \end{cases} \tag{49}$$

Similarly, we can determine the minimum required transmit power as

$$\begin{cases} p_0 = \dfrac{\gamma_0^{'} N_2^p}{|\mathbf{h}_{m0}^\dagger \bar{\mathbf{v}}|^2 - \gamma_0^{'} \|\bar{\mathbf{H}}_0 \bar{\mathbf{v}}\|^2} \\ p_m = \gamma_m \dfrac{N_m + p_0 \|\bar{\mathbf{H}}_m \bar{\mathbf{v}}\|^2}{|\mathbf{h}_{mm}^\dagger \bar{\mathbf{w}}_m|^2}, \forall m. \end{cases} \tag{50}$$

## D. Comparison for a Special Case: One CRS, One CBS and One CU

**Theorem 3** *When $M = 1$ and $r \leq 0.5$ bps/Hz, PZF always uses less power than CZF.*

*Proof:* When $M = 1$, CZF yields

$$\begin{cases} p_0 = \dfrac{\gamma_0^{'} N_2^p}{\|\mathbf{h}_{10}\|^2 (1-\rho^2)(\|\mathbf{g}_1\|^2 - \gamma_0^{'} N_{s1}) \|\mathbf{g}_1\|^2}, \\ p_1 = \dfrac{\gamma_1 N_1}{|\mathbf{h}_1^\dagger \bar{\mathbf{w}}_1|^2} = \dfrac{\gamma_1 N_1}{\|\mathbf{h}_{11}\|^2 (1-\rho^2)}, \end{cases} \tag{51}$$

where $\rho = \frac{|\mathbf{h}_{10}^\dagger \mathbf{h}_{11}|}{\|\mathbf{h}_{10}\| \|\mathbf{h}_{11}\|}$. The total power can be found as

$$P_T^{\mathsf{CZF}} = \frac{\gamma_1 N_1}{\|\mathbf{h}_{11}\|^2 (1-\rho^2)} + \frac{\gamma_0^{'} N_2^p (P_0 \|\mathbf{g}_1\|^2 + N_{s1})}{\|\mathbf{h}_{10}\|^2 (1-\rho^2)(\|\mathbf{g}_1\|^2 - \gamma_0^{'} N_{s1})}. \tag{52}$$

On the other hand, the PZF power solutions are reduced to

$$\begin{cases} p_0 = \dfrac{\gamma_0^{'} N_2^p}{\|\mathbf{h}_{10}\|^2 (\|\mathbf{g}_1\|^2 - \gamma_0^{'} N_{s1}) \|\mathbf{g}_1^\dagger\|^2} \\ p_1 = \gamma_1 \dfrac{p_0 \|\mathbf{g}_1^\dagger\|^2 \frac{|\mathbf{h}_{10}^\dagger \mathbf{h}_{11}|^2}{\|\mathbf{h}_{10}\|^2} (P_0 \|\mathbf{g}_1\|^2 + N_{s1}) + N_1}{\|\mathbf{h}_{11}\|^2 (1-\rho^2)} = \gamma_1 \dfrac{\rho^2 p_0 \|\mathbf{g}_1^\dagger\|^2 \|\mathbf{h}_{11}\|^2 (P_0 \|\mathbf{g}_1\|^2 + N_s) + N_1}{\|\mathbf{h}_{11}\|^2 (1-\rho^2)}. \end{cases} \tag{53}$$

Thus, the total power for the case of PZF is found as

$$
\begin{aligned}
P_T^{\mathsf{PZF}} &= \frac{\gamma_1 \frac{\gamma_0' N_2^p}{\|\mathbf{h}_{10}\|^2(\|\mathbf{g}_1\|^2 - \gamma_0' N_{s1})}\rho^2\|\mathbf{h}_{11}\|^2(P_0\|\mathbf{g}_1\|^2 + N_{s1}) + \gamma_1 N_1}{\|\mathbf{h}_{11}\|^2(1-\rho^2)} + \frac{\gamma_0' N_2^p(1 + P_0\|\mathbf{g}_1\|^2)}{\|\mathbf{h}_{10}\|^2(\|\mathbf{g}\|^2 - \gamma_0' N_{s1})} \\
&= \frac{\gamma_1 \gamma_0' N_2^p \rho^2(P_0\|\mathbf{g}_1\|^2 + N_{s1})}{\|\mathbf{h}_{10}\|^2(\|\mathbf{g}_1\|^2 - \gamma_0' N_{s1})(1-\rho^2)} + \frac{\gamma_1 N_1}{\|\mathbf{h}_{11}\|^2(1-\rho^2)} + \frac{\gamma_0' N_2^p(P_0\|\mathbf{g}_1\|^2 + N_{s1})}{\|\mathbf{h}_{10}\|^2(\|\mathbf{g}_1\|^2 - \gamma_0' N_{s1})}.
\end{aligned} \tag{54}
$$

As a result, we get

$$
\begin{aligned}
&P_T^{\mathsf{CZF}} - P_T^{\mathsf{PZF}} \\
&= \frac{(P_0\|\mathbf{g}_1\|^2 + N_{s1})\gamma_0' N_2^p}{\|\mathbf{h}_{10}\|^2(\|\mathbf{g}_1\|^2 - \gamma_0' N_{s1})}\left(\frac{1}{1-\rho^2} - 1 - \frac{\gamma_1\rho^2}{1-\rho^2}\right) = \frac{\gamma_0' N_2^p(P_0\|\mathbf{g}_1\|^2 + N_{s1})}{\|\mathbf{h}_{10}\|^2(\|\mathbf{g}\|^2 - \gamma_0')}\frac{\rho^2}{1-\rho^2}(1-\gamma_1),
\end{aligned} \tag{55}
$$

which indicates that PZF uses less power if and only if $\gamma_1 \leq 1$, or $r_1 = \frac{1}{2}\log_2(1+\gamma_1) \leq 0.5$ bps/Hz. $\square$

# VI. Simulation Results

Computer simulations are conducted to evaluate the performance of the proposed algorithms. We assume that the PBS has $K = 2$ transmit antennas and there are $M = 3$ CBSs, each with $N = 4$ antennas, serving 3 CUs. The channel between any antenna pair is modeled as $h = d^{-\frac{c}{2}}e^{j\theta}$, where $d$ is the distance, $c$ is the path loss exponent, chosen as 3.5, and $\theta$ is uniformly distributed over $[0, 2\pi)$. The distances from the CBSs to the PBS, PU and CUs are all normalized to one while the distance from the PBS to the PU is set to 2 units. Thus, the primary channel is much weaker than other links. The PU's target rate is $r_0 = 2$ bps/Hz unless otherwise specified. For convenience, all the CUs have the same target rate $r$ and noise power levels at all terminals are the same. We define the transmit SNR, transmit power normalized by noise power, as the power metric. For each result, $10^4$ channel realizations are simulated and averaged. Outage is said to occur when $r_0$ is not supported in the primary system for a channel instance and in this case the primary system will have cooperation with the CBS. When outage does not occur even without cooperation, the CBS operates in Mode II.2, i.e., it transmits to the CUs in the orthogonal space of the PU.

We study the convergence behaviors of Algorithms 1 and 2 in Fig. 3 for a typical channel realization assuming $r = 1$ bps/Hz. For fairness, both algorithms start with a zero power vector initialization. The relative change in the total power is shown as the iteration goes. Results illustrate that 5 iterations are sufficient for Algorithm 2 to converge while Algorithm 1 converges much slower. Fig. 3 also reveals that Algorithm 1 and Algorithm 2 have linear and superlinear convergence rates, respectively. This verifies our theoretical analysis that Algorithm 2 converges significantly faster than Algorithm 1.

The impact of PBS power on the required total CBS transmit power is studied in Fig. 4 assuming that each CU requires a minimum rate of 2 bps/Hz. Results show that the optimal solution greatly outperforms both zero-forcing solutions and saves about 3 dB in transmit SNR for the CBSs when the transmit SNR for the PBS increases from 4 to 12 dB. Also, PZF uses about 4 and 10 dB more power than CZF and the optimal solution, respectively. This is because PZF gives priority to the PU and compromises the performance of CUs, but for CZF, the CUs and the PU are equally important.

Fig. 5 demonstrates the cumulative density function (CDF) of the CBS transmit SNR against the CU target rates, assuming that the PBS transmit SNR is 10 dB. As we can see, when $r = 0.1$ bps/Hz, PZF outperforms CZF because the PU's rate constraint dominates, thereby giving priority to the PU. When $r = 0.5$ bps/Hz, CZF and PZF have almost identical performance, whereas when $r > 0.5$ bps/Hz, CZF greatly outperforms PZF as PZF gives priority to the PU and compromises the performance of CUs.

Next we examine the effect of the CBS power on the PU's outage performance in Fig. 6. Results for "No Cooperation" in which the PU uses the entire available bandwidth for transmission without letting the CBSs to use part of the bandwidth for cooperation is provided for comparison. We consider that the CUs have a target rate of 2 bps/Hz and the PBS's transmit SNR is 10 dB. Results reveal that outage occurs with a probability of 85% for the PU but the situation greatly improves when the PBS cooperates with the CBSs and if the CBS transmit power is significant. At low to medium CBS transmit SNR (0–15 dB), cooperation does not pay off (but also does not worsen the PU outage performance) but as the CBS transmit power increases beyond 15 dB, the PU's outage probability drops rapidly due to the increased diversity from the CBSs, showing that cooperation is a promising approach to improve the PU link.

Finally, we evaluate the robustness of the proposed algorithms against imperfect CSI knowledge between CBSs and the PU. To model the errors, we assume that

$$\mathbf{h}_{j0} = \hat{\mathbf{h}}_{j0} + \Delta\mathbf{h}_{j0}, \forall j, \tag{56}$$

where $\hat{\mathbf{h}}_{j0}$ is the CSI known at CBS $j$ and $\Delta\mathbf{h}_{j0} \sim \mathcal{CN}(0, \xi_j^2\mathbf{I})$ denotes the CSI error. We consider a system with one CBS and one CU and the PBS's transmit SNR is 10 dB. Fig. 7 shows the achievable PU rates (with a target of 2 bps/Hz) for the three proposed schemes against $\xi^2$. It is seen that the optimal solution and CZF solution are quite robust while PZF is very sensitive to the error. This can be explained that in PZF, CBSs choose beamforming vectors aligned with $\{\hat{\mathbf{h}}_{j0}\}$, thereby suffering most. Another important observation is that the achievable rate for the PU is worst without cooperation and even at high level of CSI errors, cooperation gains and the CZF appears to be a promising cooperation strategy.

# VII. Conclusion

We addressed the jointly optimization problem of collaborative relaying at CBSs to relay the primary signals, and coordinated transmit beamforming for cognitive transmission. The problem of minimizing the CBS transmit power subject to minimum rate requirements for the PU and the CUs was studied. The optimal structure of the relaying matrix was shown to first match the backward channels and then retransmit the noisy PU signal using collaborative relay beamforming. Two efficient algorithms to find the optimal solution were proposed and compared. A novel distributed implementation was also devised for the cooperation. Also, suboptimal but closed-form CZF and PZF solutions were presented.

# Appendix I. Proof of Theorem 1

Our derivation is based on the analysis of KKT conditions. To do so, we write the Lagrangian of (7) as

$$\begin{aligned}\mathcal{L} = & \sum_{j=1}^{M} \|\mathbf{w}_j\|^2 + P_0 \sum_{j=1}^{M} \|\mathbf{A}_j \mathbf{g}_j\|^2 + \sum_{j=1}^{M} N_{sj} \|\mathbf{A}_j\|^2 \\ & + \lambda_0 \left( \sum_{j=1}^{M} |\mathbf{h}_{j0}^{\dagger} \mathbf{w}_j|^2 + \sum_{j=1}^{M} N_{sj} \|\mathbf{h}_{j0}^{\dagger} \mathbf{A}_j\|^2 + N_2^p - \frac{|\sum_{j=1}^{M} \mathbf{h}_{j0}^{\dagger} \mathbf{A}_j \mathbf{g}_j|^2}{\gamma_0^{'}} \right) \\ & + \sum_{m=1}^{M} \lambda_m \left( \sum_{\substack{j=1 \\ j \neq m}}^{M} |\mathbf{h}_{jm}^{\dagger} \mathbf{w}_j|^2 + P_0 \sum_{j=1}^{M} |\mathbf{h}_{jm}^{\dagger} \mathbf{A}_j \mathbf{g}_j|^2 + \sum_{j=1}^{M} N_{sj} \|\mathbf{h}_{jm}^{\dagger} \mathbf{A}_j\|^2 + N_m - \frac{|\mathbf{h}_{mm}^{\dagger} \mathbf{w}_m|^2}{\gamma_m} \right), \end{aligned} \quad (57)$$

where $\lambda_0, \{\lambda_m\}$ are dual variables. Setting $\frac{\partial \mathcal{L}}{\partial \mathbf{A}_j^*} = \mathbf{0}$ leads to

$$\begin{aligned} & P_0 \mathbf{A}_j \mathbf{g}_j \mathbf{g}_j^{\dagger} + N_{sj} \mathbf{A}_j + \lambda_0 N_{sj} \mathbf{h}_{j0} \mathbf{h}_{j0}^{\dagger} \mathbf{A}_j + P_0 \sum_{m=1}^{M} \lambda_m \mathbf{h}_{jm} \mathbf{h}_{jm}^{\dagger} \mathbf{A}_j \mathbf{g}_j \mathbf{g}_j^{\dagger} + \sum_{m=1}^{M} \lambda_m N_s j \mathbf{h}_{jm} \mathbf{h}_{jm}^{\dagger} \mathbf{A}_j \\ & = \frac{\lambda_0}{\gamma_0^{'}} \mathbf{h}_{j0} \mathbf{h}_{j0}^{\dagger} \mathbf{A}_j \mathbf{g}_j \mathbf{g}_j^{\dagger} + \frac{\lambda_0}{\gamma_0^{'}} \sum_{\substack{n=1 \\ n \neq j}}^{M} \mathbf{h}_{j0} \mathbf{h}_{n0}^{\dagger} \mathbf{A}_n \mathbf{g}_j \mathbf{g}_j^{\dagger}, \end{aligned} \quad (58)$$

where we have used $\frac{\partial \mathsf{trace}(\mathbf{A}\mathbf{T}_1\mathbf{A}^\dagger\mathbf{T}_2)}{\partial \mathbf{A}^*} = \mathbf{T}_2\mathbf{A}\mathbf{T}_1$. Re-organizing (58), we obtain

$$N_{sj}\left(\mathbf{I}+\lambda_0\mathbf{h}_{j0}\mathbf{h}_{j0}^\dagger+\sum_{m=1}^{M}\lambda_m\mathbf{h}_{jm}\mathbf{h}_{jm}^\dagger\right)\mathbf{A}_j+P_0\left(\mathbf{I}+\lambda_0\mathbf{h}_{j0}\mathbf{h}_{j0}^\dagger+\sum_{m=1}^{M}\lambda_m\mathbf{h}_{jm}\mathbf{h}_{jm}^\dagger\right)\mathbf{A}_j\mathbf{g}_j\mathbf{g}_j^\dagger$$
$$=\frac{\lambda_0}{\gamma_0'}\mathbf{h}_{j0}\mathbf{h}_{j0}^\dagger\mathbf{A}_j\mathbf{g}_j\mathbf{g}_j^\dagger+\lambda_0\mathbf{h}_{j0}\mathbf{h}_{j0}^\dagger\mathbf{A}_j\mathbf{g}_j\mathbf{g}_j^\dagger+\frac{\lambda_0}{\gamma_0'}\sum_{\substack{n=1\\n\neq j}}^{M}\mathbf{h}_{j0}\mathbf{h}_{n0}^\dagger\mathbf{A}_n\mathbf{g}_j\mathbf{g}_j^\dagger\propto\mathbf{h}_{j0}\mathbf{g}_j^\dagger$$
$$\Rightarrow\mathbf{A}_j\left(N_{sj}\mathbf{I}+P_0\mathbf{g}_j\mathbf{g}_j^\dagger\right)\propto\left(\mathbf{I}+\lambda_0\mathbf{h}_{j0}\mathbf{h}_{j0}^\dagger+\sum_{m=1}^{M}\lambda_m\mathbf{h}_{jm}\mathbf{h}_{jm}^\dagger\right)^{-1}\mathbf{h}_{j0}\mathbf{g}_j^\dagger. \quad (59)$$

Applying the matrix inversion lemma to (59), we have the result

$$\left(\mathbf{I}+\lambda_0\mathbf{h}_{j0}\mathbf{h}_{j0}^\dagger+\sum_{m=1}^{M}\lambda_m\mathbf{h}_{jm}\mathbf{h}_{jm}^\dagger\right)^{-1}\mathbf{h}_{j0}\mathbf{g}_j^\dagger$$
$$\propto\left(\mathbf{I}+\sum_{m=1}^{M}\lambda_m\mathbf{h}_{jm}\mathbf{h}_{jm}^\dagger\right)^{-1}\mathbf{h}_{j0}\mathbf{g}_j^\dagger\propto\left(a_{j0}\mathbf{h}_{j0}+\sum_{m=1}^{M}a_{jm}\mathbf{h}_{jm}\right)\mathbf{g}_j^\dagger. \quad (60)$$

Combining (59) and (60) gives

$$\mathbf{A}_j\propto\left(a_{j0}\mathbf{h}_{j0}+\sum_{m=1}^{M}a_{jm}\mathbf{h}_{jm}\right)\mathbf{g}_j^\dagger\left(N_{sj}\mathbf{I}+P_0\mathbf{g}_j\mathbf{g}_j^\dagger\right)^{-1}\propto\left(a_{j0}\mathbf{h}_{j0}+\sum_{m=1}^{M}a_{jm}\mathbf{h}_{jm}\right)\mathbf{g}_j^\dagger, \quad (61)$$

where we have used the matrix inversion lemma and $a_{j0}, \{a_{jm}\}$ are all (complex) constants.

Now, we use a composite channel matrix $\mathbf{H}_j = [\mathbf{h}_{j0}\ \mathbf{h}_{j1}\cdots\mathbf{h}_{jM}]$ from $\mathsf{CBS}_j$ to all the PU and CUs, and introduce a new vector $\mathbf{a}_j = [a_{j0}\ a_{j1}\cdots a_{jM}]^T$. Then, we reach the optimal structure in the theorem.

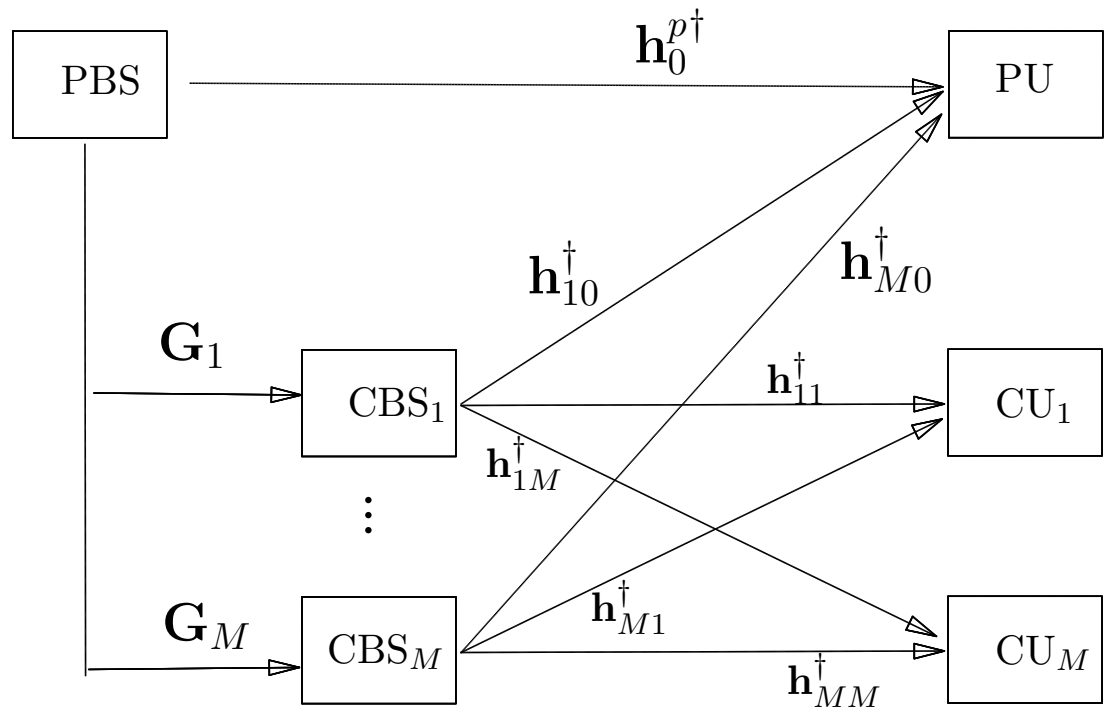

Figure 1: The cooperating cognitive system.

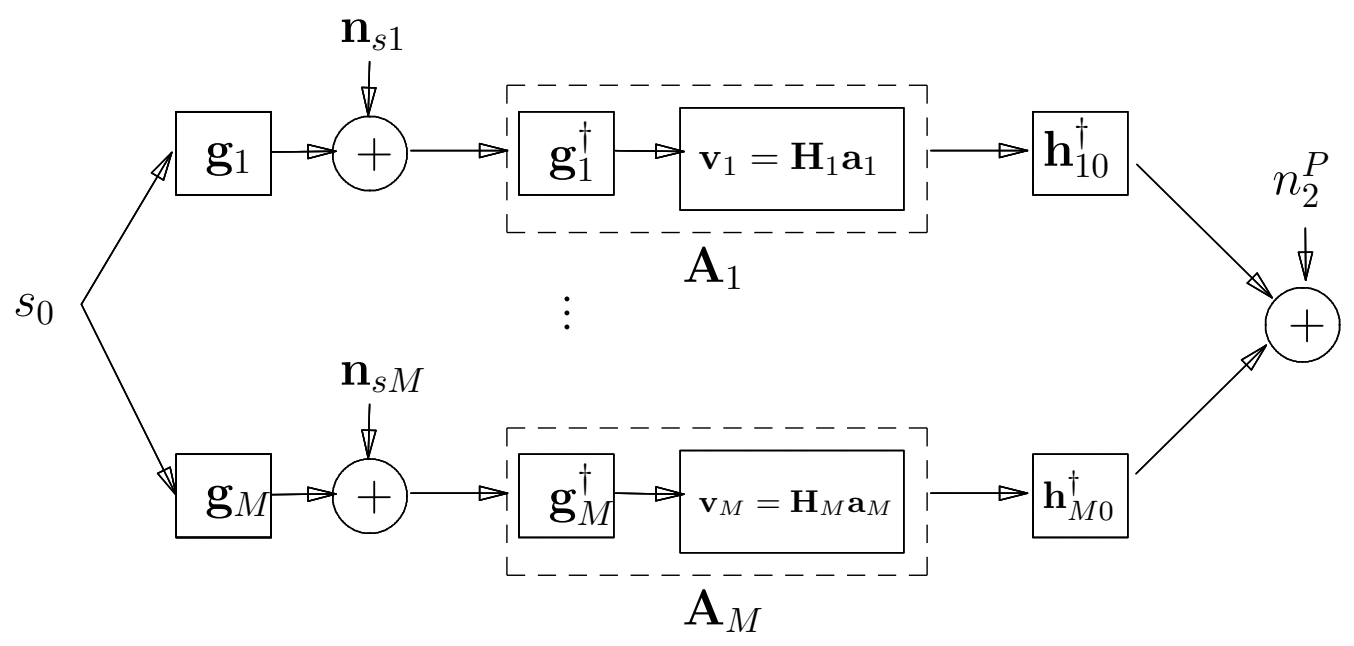

Figure 2: Illustration of the relaying processing at the CBSs.

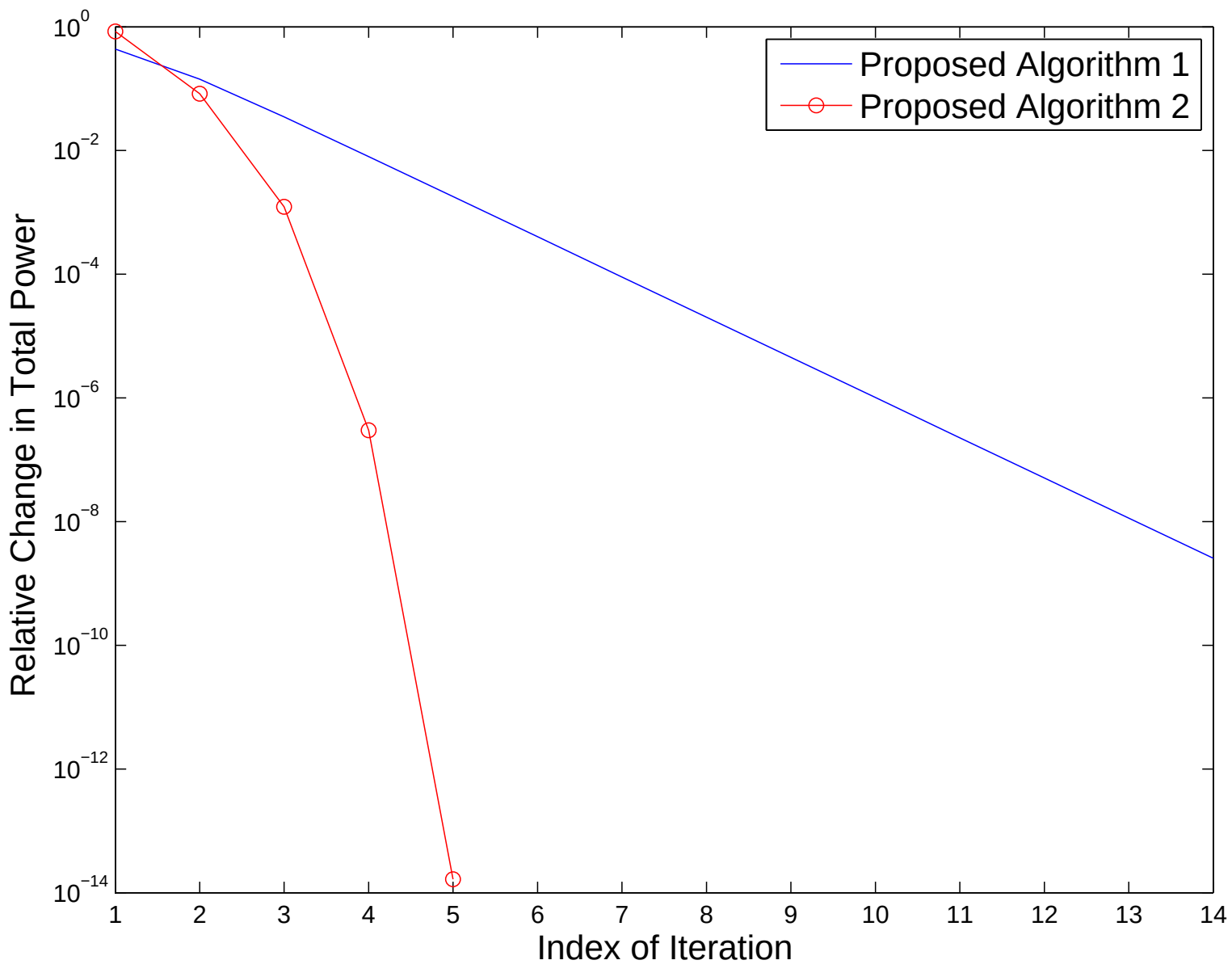

Figure 3: The convergence behavior of the proposed algorithms.

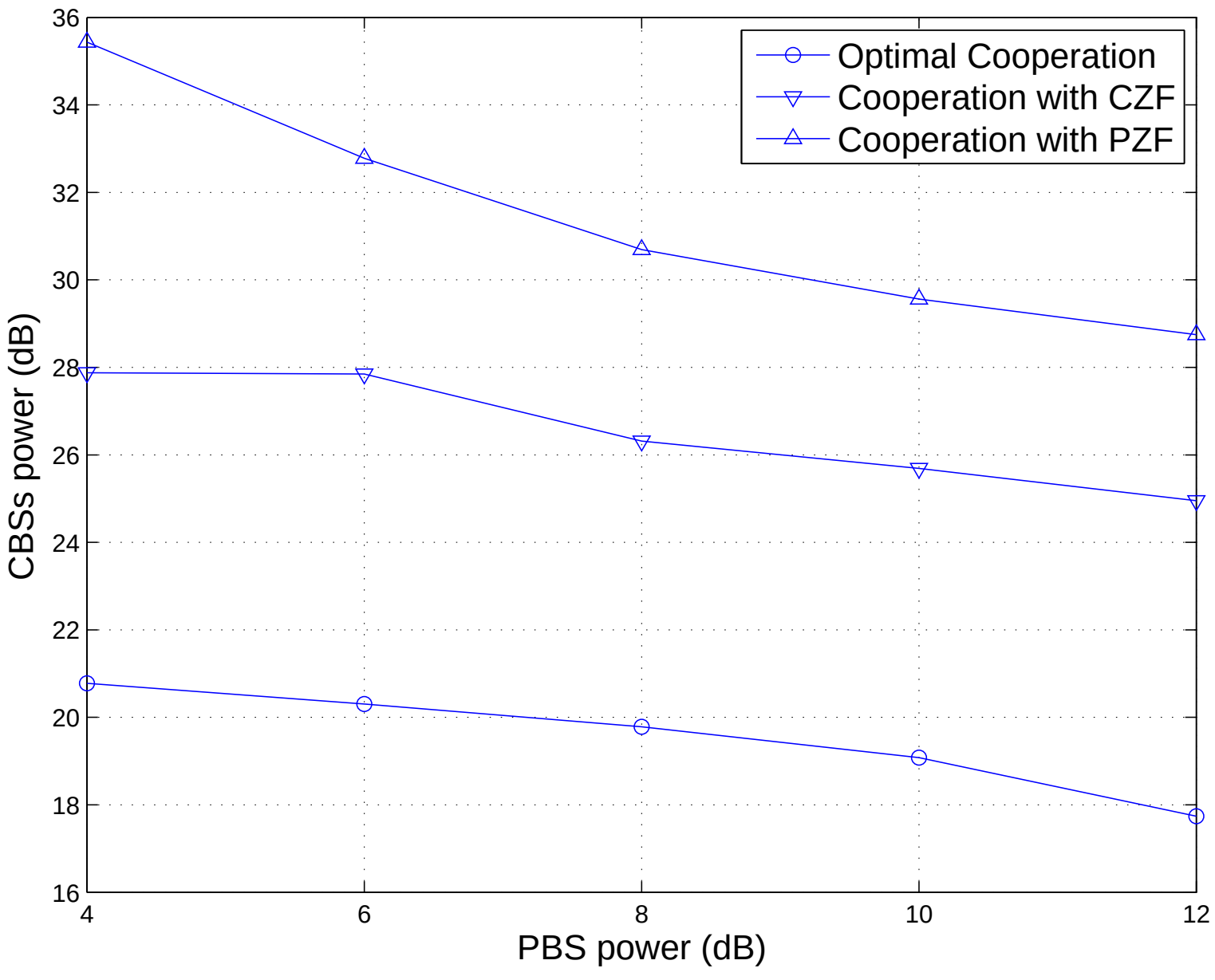

Figure 4: The required CU power versus the PBS power.

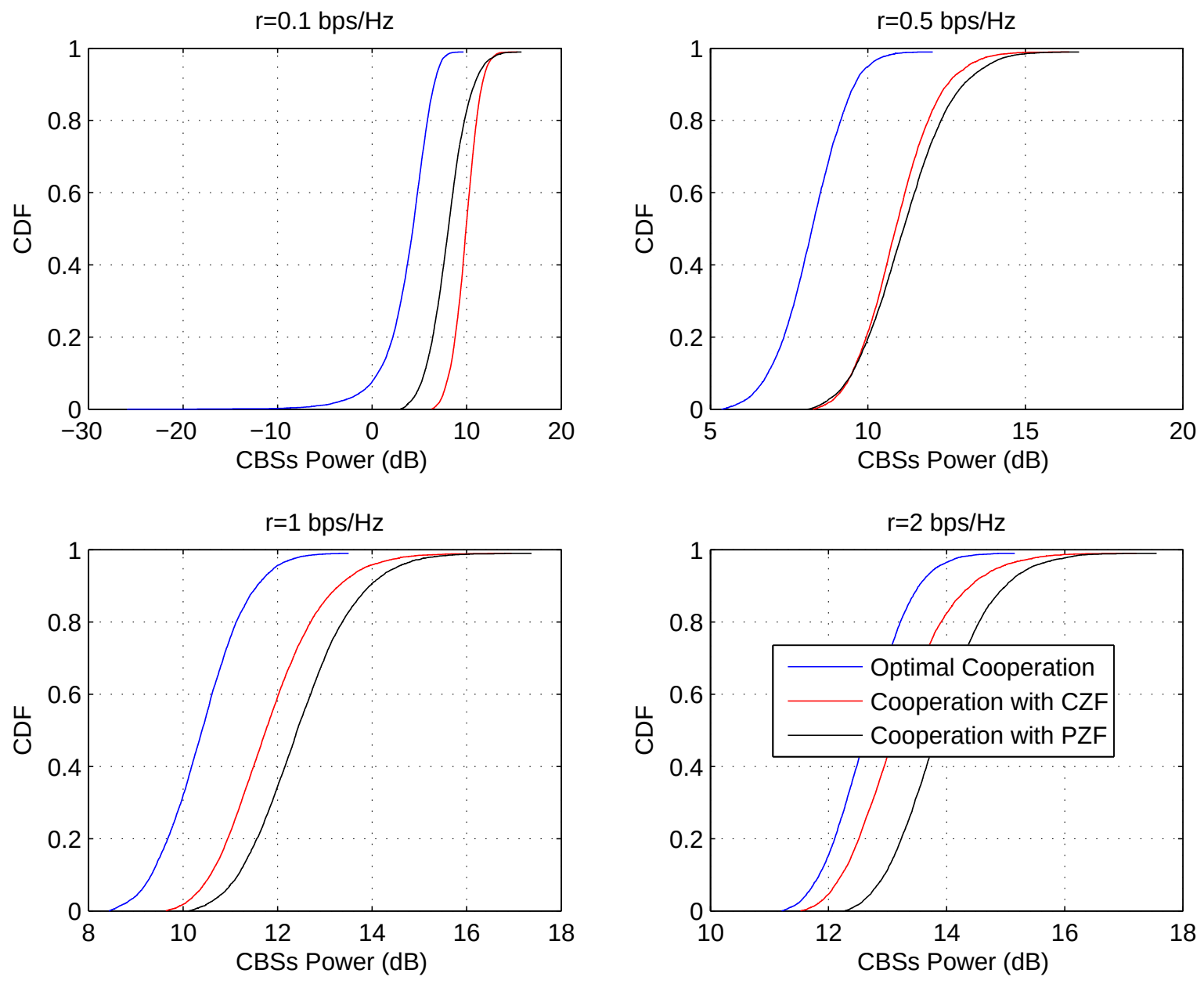

Figure 5: The CDF results of the average CBS power.

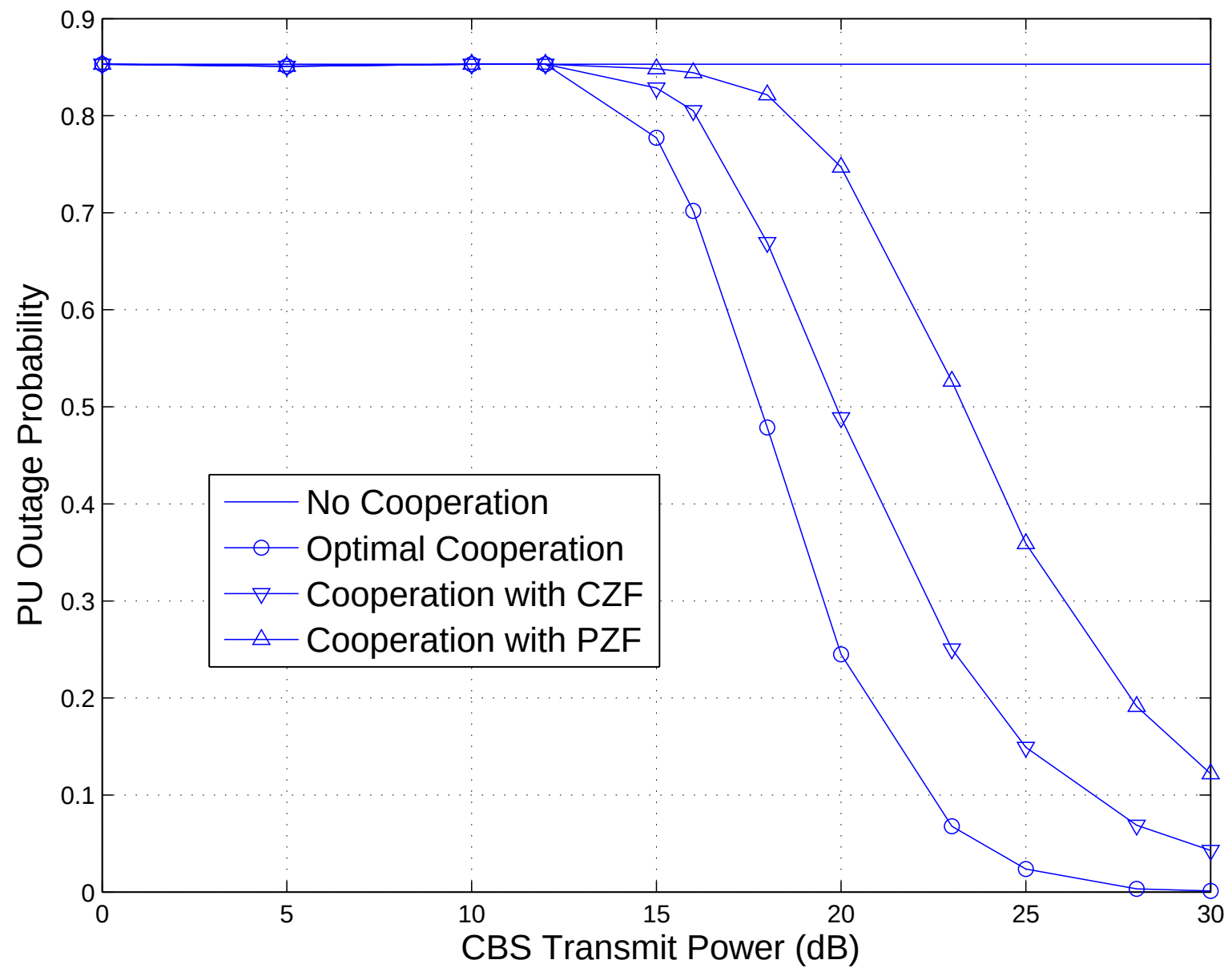

Figure 6: The PU outage probability against the CBS power.

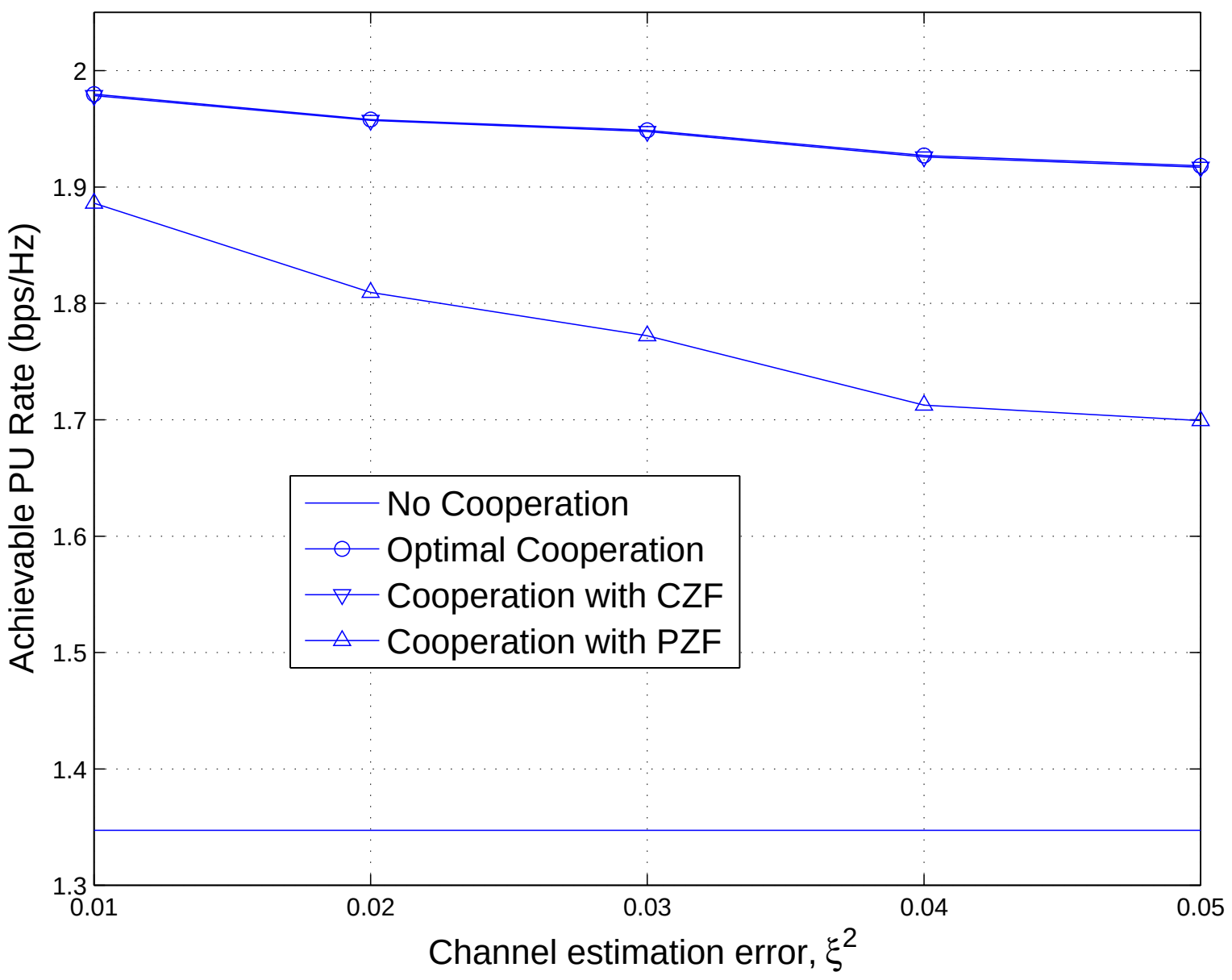

Figure 7: The PU rate against the channel error.